\documentclass[letterpaper,aps,prc,superscriptaddress,nofootinbib,showpacs,floatfix,10pt]{revtex4-1}
\usepackage{hyperref}
\usepackage[utf8]{inputenc}
\usepackage{amsmath}
\usepackage{amsfonts}
\usepackage{amssymb}
\usepackage{amsthm}
\usepackage{bm}
\usepackage[english]{babel}
\usepackage{fontenc}
\usepackage{graphicx}
\usepackage[margin=1in]{geometry}
\usepackage{enumerate}
\usepackage{textcomp}
\usepackage[toc,page]{appendix}
\usepackage{slashed}
\usepackage{color}
\usepackage{MnSymbol}
\usepackage{setspace}
\graphicspath{ {figure/}}
\hypersetup{colorlinks=true, linkcolor=blue, citecolor=blue, urlcolor=blue}

{\large }

\begin{document}
\title{Multi-particle Integral and Differential Correlation Functions}
\author{ Claude~Pruneau }
\email{claude.pruneau@wayne.edu}
\affiliation{Department of Physics and Astronomy, Wayne State University, Detroit, 48201, USA}
\author{ Victor~Gonzalez }
\email{victor.gonzalez@cern.ch}
\affiliation{Department of Physics and Astronomy, Wayne State University, Detroit, 48201, USA}
\author{Ana Marin} 
\email{a.marin@gsi.de}
\affiliation{GSI Helmholtzzentrum f\"ur Schwerionenforschung GmbH, Research Division and ExtreMe Matter Institute EMMI, Darmstadt, Germany}
\author{ Sumit~Basu }
\email{sumit.basu@cern.ch}
\affiliation{Lund University, Department of Physics, Division of Particle Physics, Box 118, SE-221 00, Lund, Sweden}

\begin{abstract}
This paper formalizes the use of integral and differential cumulants for measurements of multi-particle event-by-event transverse momentum fluctuations, rapidity fluctuations, as well as net charge fluctuations. This enables the introduction of multi-particle balance functions, defined based on differential correlation functions (factorial cumulants), that  suppress two and three prong resonance decays effects and enable measurements of underlying long range correlations obeying quantum number conservation constraints. 
These multi-particle balance functions satisfy  simple sum rules determined by quantum number conservation.  It is additionally shown that these multi-particle balance functions arise as an intrinsic component of high-order net charge cumulants. This implies that the magnitude of these cumulants, measured in a specific experimental acceptance,  is strictly constrained by  charge conservation and primarily determined by the rapidity and momentum width of these 
balance functions.
The paper also presents techniques to reduce the computation time of differential correlation functions up to order $n=$10 based on the methods of moments.

\end{abstract}

\maketitle

\section{Introduction}

A variety of two-particle integral and  differential  correlation functions have been developed and deployed towards the analysis of particle production in heavy-ion collisions~\cite{STAR:2003kbb,Agakishiev:2011fs,ALICE:2018jco, Adam:2017ucq}. Correlation functions formulated as functions of rapidity and azimuth angle differences have enabled, in particular, the discovery of away-side jet suppression in large collision systems~\cite{RHICqgp1,RHICqgp2}. Subsequent measurements of long range particle correlations in both large and small collision systems have additionally enabled detailed studies of the collision dynamics. Of particular interest are two-particle number correlations, $R_2$, based on normalized two-particle differential cumulants~\cite{Adam:2017ucq,ALICE:2018jco}, transverse momentum ($p_{\rm T}$) correlation functions, $P_2$, designed to study $p_{\rm T}$ fluctuations~\cite{Sharma:2008qr}, and $G_2$, designed for the study of viscous effects based on their longitudinal broadening in A--A collisions~\cite{STAR:2011iio,ALICE:2019smr,ALICE:2022hor}. Differential two-particle correlation functions have been studied for charge inclusive (CI) particle pairs as well as charge dependent (CD) pairs. The latter are of particular interest because they relate to charge balance functions, which provide a tool to study the correlation length, in momentum space, of charge balancing particles. Balance functions were initially developed to investigate the presence of isentropic expansion and delayed hadronization in large collision systems~\cite{Bass:2000az,Jeon:2002BFCF,Basu:2020ldt}. However, balance functions were later  shown to also be sensitive to quark susceptibilities~\cite{PhysRevC.99.044916} as well as the diffusivity of light quarks~\cite{Pratt:2021xvg,Pratt:2019pnd}. Recent works have shown they are best measured in terms of differential two-particle cumulants and feature simple sum-rule that might be instrumental in the study of thermal hadron production~\cite{Pruneau:2019baa,Pruneau:2022brh,Pruneau:2023zhl}.

Two-particle correlation functions are by construction dominated by  contributions from  two particle correlated production processes such as hadronic decays and  jet production. They are also nominally sensitive to higher order particle correlations even though they cannot specifically discriminate such higher order processes. In many instances, it is these higher order correlation processes that are of interest to investigate particle production and properties of the matter produced in A--A collisions. As a first example, consider measurements of transverse momentum fluctuation deviates. Correlators of transverse momentum deviates were initially invoked as a proxy to study temperature fluctuations in A--A collisions~\cite{Voloshin:1999yf}. However,  it may be argued that fluctuations observed in A--A collisions are strictly driven by initial state conditions~\cite{Gardim:2020sma}. There is thus interest in establishing the magnitude of such collective fluctuations. The use of the multiple particle $p_{\rm T}$ integral correlator has then been advocated to study the effects of initial fluctuations~\cite{Giacalone:2020lbm,ALICE:2023tej}. However note that ongoing studies  may suffer from short-range correlations (a.k.a. non-flow) and it might  thus be  desirable to develop differential analysis techniques that enable rapidity gaps designed to suppress short range correlations. As a second example of interest, consider charge, strangeness, and baryon balance functions. In this case also, one expects the correlations to receive sizable contributions from two and three prong particle decays as well as short range correlation from jets. There is thus an interest in obtaining higher order correlation functions that are sensitive to charge (strangeness, baryon) balance but suppress contributions from decays and jets and it is the primary purpose of this work to develop 
such multi-particle balance functions.  As in studies of anisoptropic flow measurements,  multi-particle cumulants, of order $n$=4, 6, $\ldots$, shall be used to suppress lower order correlations and obtain  multi-particle balance functions.  Unfortunately, measurements and  calculations of higher order cumulants nominally require several  nested loops to include contributions from all $n$-tuplets of interest on an event-by-event basis. Although such calculations are conceptually simple and remain practical for collisions and experimental acceptances featuring a modest particle multiplicity, they become prohibitively CPU intensive for large  multiplicities. 
Fortunately, a number of techniques have been developed, particularly in the context of anisotropic flow analyses, to reduce the computational challenge of handling large multiplicities and high-order cumulants~\cite{Bilandzic:2013kga}. One such application successfully deployed in the context of anisotropic flow studies and towards the computation of higher moment deviates relies of the method of moments~\cite{Giacalone:2020lbm}. A second purpose of this paper is to further develop this method towards measurements of differential multi-particle correlations of inclusive as well as identified particle species. 

This paper is organized as follows. First, sec.~\ref{sec:motivation} presents a  discussion detailing the need for higher order  integral and differential correlators based on several  examples. Higher order correlators are 
then introduced in sec.~\ref{sec:multiparticle} in terms of integral and differential cumulants of arbitrary order as well as  expectation values of the form $\llangle q_1 q_2 \cdots q_n\rrangle$ and $\llangle \Delta q_1 \Delta q_2 \cdots \Delta q_n\rrangle$, where $q_i$, $i=1,\ldots, n$ represent particle variables of interest (e.g., transverse momentum, charge, etc) and  $\Delta q_i=q_i -\llangle q\rrangle$, $i=1,\ldots, n$, are their deviates relative to their event ensemble average $\llangle q\rrangle$. Techniques to compute these expectation values  based on the method of moments are presented in secs.~\ref{sec:method} and \ref{sec:configurations}. 
Equipped with these different correlators and computing tools, the notion of multi-particle balance function is then introduced in sec.~\ref{sec:Balance Functions}.  
Finally, multi-particle balance functions are  considered in sec.~\ref{sec:Net-ChargeCumulants} in the context of measurements of net-charge cumulants and it is shown that net charge cumulants of all orders are explicitly constrained by charge conservation. The paper is summarized in sec.~\ref{sec:summary}.
Given much of the calculations performed  in this work are somewhat tedious and lengthy, 
details of these calculations  are presented in several appendices. Appendix~\ref{sec:MomentsCumulants} presents calculation methods and formula for moments, cumulants, factorial moments, and factorial cumulants for single and multi-variable systems, as well as for the net-charge of particles measured in a specific acceptance $\Omega$. Appendix~\ref{sec:DifferentialCorrelations} extends these formula to differential correlations. General formula for the computations of deviates of the form $\llangle \Delta q_1 \Delta q_2 \, \cdots \, \Delta q_m\rrangle$ are derived in 
appendix~\ref{sec:DeviatesAverage} while equations for the calculation of correlators 
of the form $\llangle  q_1   q_2 \, \cdots \,  q_m p_1  \, \cdots \,  p_n \rrangle$ are listed in appendix.~\ref{sec:computation}. 
Finally, appendix~\ref{sec:Multi-particleBalanceFunctions} lists definitions of multi-particle balance functions up to order $n=10$.

\section{Motivations
\label{sec:motivation}}
Let us consider single particle densities of particles of type $\alpha$, denoted $\rho_1^{\alpha}(\vec p)$, and  $n$-particle densities of mixed species $\alpha_1, \ldots, \alpha_n$, denoted $\rho_n^{\alpha_1\cdots\alpha_n}(\vec p_1,\ldots,\vec p_n)$. In general, mixed $n$-particle densities, $\rho_n^{\alpha_1\cdots\alpha_n}(\vec p_1,\ldots,\vec p_n)$, correspond to  the yield (per event) of $n$-tuplets of particles of types $\alpha_1, \alpha_2, \ldots, \alpha_n$ at momenta $\vec p_1,\ldots,\vec p_n$. Such $n$-tuplets may arise from a single process yielding $n$ correlated (mixed) particles, or  combinations of processes jointly yielding $n$-particles. In order to focus a study on correlated particles exclusively, one commonly relies on the notions of integral and differential correlation function cumulants. Indeed recall that, by construction, integration of $\rho_1^{\alpha}(\vec p)$ over a specific kinematic range  $\Omega$  yields the average  number of  particles of type $\alpha$ in this acceptance, whereas integration of two-, three-, or $n$ mixed particle densities yield the average number of pairs, triplets, and more generally $n$-tuplets of such groupings of particles.
These integrals do not discriminate correlated from uncorrelated particles.

Differential cumulants and their integrals are of particular interest because they identically  vanish  in the {\it absence} of $n$ or more particle correlations. They are thus an essential tool for the study of particle production. They can also be straightforwardly corrected for uncorrelated particle losses (detection efficiency). 
This feature is exploited in the formulation of number correlation function ratios such as
\begin{align}
\label{eq:DiffNormCumulant2}
R_2^{\alpha\beta}(\vec p_1,\vec p_2)
&\equiv
\frac{ \rho_2^{\alpha\beta}(\vec p_1,\vec p_2) - \rho_1^{\alpha}(\vec p_1) \rho_1^{\beta}(\vec p_2)}
  { \rho_1^{\alpha}(\vec p_1) \rho_1^{\beta}(\vec p_2)}
\equiv \frac{ C_2^{\alpha\beta}(\vec p_1,\vec p_2)}
  { \rho_1^{\alpha}(\vec p_1) \rho_1^{\beta}(\vec p_2)}
\equiv \frac{F_2^{\alpha_1\alpha_2}}{F_1^{\alpha_1}F_1^{\alpha_2}}
\end{align}
with $C_{2}$ the second order cumulant and $F_{1}$ and $F_2$ the first and second order factorial
cumulants, which is said to be robust against particle losses, i.e., independent of efficiencies provided these are approximately constant within the acceptance of a measurement~\cite{Pruneau:2002yf}. Ratios of factorial cumulants, similarly formulated, have also been used in  recent studies~\cite{Agakishiev:2011fs,Adam:2017ucq}. They too feature the property of robustness against particle (efficiency) losses. Particular combinations of differential and integral ratios $R_2^{\alpha\beta}$ have also been used in the context of relative yield fluctuation studies and balance functions~\cite{Pruneau:2017ypa}.

Integral and differential correlation functions have also been used to study fluctuations of specific kinematic variables. Fluctuations of event-wise total transverse momentum, in particular, have been proposed to study temperature and energy fluctuations in the initial stage of heavy-ion collisions. Voloshin et al.~\cite{Voloshin:1999yf} showed fluctuations measures are best formulated in terms of transverse momentum deviates  $\Delta p_{\rm T,1}\Delta p_{\rm T,2}$ according to
\begin{equation}
\label{eq:DptDptInt}
\llangle \Delta p_{\rm T,1}\Delta p_{\rm T,2} \rrangle \equiv \frac{
\int_{\Omega} \Delta p_{\rm T,1}\Delta p_{\rm T,2}\hspace{0.05in} \rho_2(\vec p_1,\vec p_2)\, {\rm d}\vec p_1\,{\rm d}\vec p_2
}{
\int_{\Omega} \rho_2(\vec p_1,\vec p_2)\, {\rm d}\vec p_1\,{\rm d}\vec p_2}.
\end{equation}
This integral correlator is commonly reported in terms of a dimensionless ratio $\llangle \Delta p_{\rm T,1}\Delta p_{\rm T,2} \rrangle/\llangle p_{\rm T}\rrangle^2$~\cite{Voloshin:2002ku,Abelev:2014ckr}.
A differential version of this dimensionless correlator was also used~\cite{Sharma:2008qr,Agakishiev:2011fs,Adam:2017ucq,Sahoo:2022jgm} and can be written according to
\begin{equation}
\label{eq:DptDptDiff}
P_2(y_1,\varphi_1,y_2,\varphi_2) \equiv
\frac{1}{ \llangle p_{\rm T}\rrangle^2}
\frac{
\int_{\Omega} \Delta p_{\rm T,1}\Delta p_{\rm T,2}\hspace{0.05in} \rho_2(p_{\rm T,1}, y_1,\varphi_1,p_{\rm T,2}y_2,\varphi_2)\, {\rm d} p_{\rm T,1}{\rm d} p_{\rm T,2}
}{
\int_{\Omega} \rho_2(p_{\rm T,1}, y_1,\varphi_1,p_{\rm T,2}, y_2,\varphi_2)\,  {\rm d} p_{\rm T,1}{\rm d}p_{\rm T,2}}.
\end{equation}

A generalization of $\llangle \Delta p_{\rm T,1}\Delta p_{\rm T,2} \rrangle$ to four particle correlations was first proposed by Voloshin~\cite{Voloshin:2002ku} towards the study of temperature and energy fluctuations.
 The study of third and fourth  $p_{\rm T}$ moments,
defined according to \begin{align}
\label{eq:DptDptDptInt}
\llangle \Delta p_{\rm T,1}\Delta p_{\rm T,2}\Delta p_{\rm T,3} \rrangle =& \frac{1}{ \llangle p_{\rm T}\rrangle^3}
\frac{
\int_{\Omega} \Delta p_{\rm T,1}\Delta p_{\rm T,2}\Delta p_{\rm T,3} \hspace{0.05in}\rho_3(\vec p_1,\vec p_2,\vec p_3) \,{\rm d}\vec p_1\,{\rm d}\vec p_2\,{\rm d}\vec p_3}
{\int_{\Omega} \rho_3(\vec p_1,\vec p_2,\vec p_3) \,{\rm d}\vec p_1\,{\rm d}\vec p_2\,{\rm d}\vec p_3} \\
\label{eq:DptDptDptDptInt}
\llangle \Delta p_{\rm T,1}\Delta p_{\rm T,2}\Delta p_{\rm T,3}\Delta p_{\rm T,4} \rrangle  =&  \frac{1}{ \llangle p_{\rm T}\rrangle^4}
\frac{
\int_{\Omega} \Delta p_{\rm T,1}\Delta p_{\rm T,2}\Delta p_{\rm T,3} \Delta p_{\rm T,4}\hspace{0.05in} \rho_4(\vec p_1,\vec p_2,\vec p_3,\vec p_4) \,{\rm d}\vec p_1
\,{\rm d}\vec p_2 \,{\rm d}\vec p_3 \,{\rm d}\vec p_4}
{\int_{\Omega} \rho_4(\vec p_1,\vec p_2,\vec p_3,\vec p_4) \,{\rm d}\vec p_1
\,{\rm d}\vec p_2 \,{\rm d}\vec p_3 \,{\rm d}\vec p_4},
\end{align}
was  proposed to probe initial stage fluctuations~\cite{Giacalone:2020lbm}.  Measurements of these correlators were recently reported by the ALICE collaboration~\cite{ALICE:2023tej}. Clearly, it is trivial to also consider differential versions of these two correlators and such generalizations of   $P_2(y_1,\varphi_1,y_2,\varphi_2)$ might be useful to carry out higher moment analyses with finite rapidity gaps. Additionally, as we discuss in the next sections, extension to particle correlators of this form to $n>4$ particles are readily accessible based on the methods of moments presented in sec.~\ref{sec:method}.

The integral and differential correlation functions, Eqs.~(\ref{eq:DptDptInt}--\ref{eq:DptDptDptDptInt}), may also be
applicable to the study of other types of fluctuations. For instance, replacing $\Delta p_{\rm T,i}$ by rapidity deviates
$\Delta y_{i}\equiv y_i -\llangle y \rrangle$, where $y_i$ are the rapidities of particles $i=1, \ldots, N$ of an event, it becomes possible to study event-by-event fluctuations in the rapidity of particles. Such fluctuations might provide an alternative way to probe the longitudinal  correlation length (rapidity) of produced particles.

The multi-particle correlators and the set of tools for their extraction presented in Sec.~\ref{sec:multiparticle} provide a basis for the  extension of former techniques used for measurements of transverse momentum correlations, rapidity correlations, as well as charge  correlations (including baryon and strangeness numbers correlations) heretofore completed mostly at low orders $n\le 4$. These techniques also connect to measurements of anisotropic flow and correlations between flow and other variables discussed elsewhere~\cite{Bilandzic:2013kga}.

The techniques, whether used with a single or several variables, are nominally very powerful because they enable joint measurements involving many particles simultaneously. However, it is also clear that the complexity of such measurements can quickly grow out of hand. For instance, assuming an interest in the rapidity, transverse momentum, and azimuth of particles, one would nominally get differential observables $\llangle \Delta q_1 \Delta q_2 \cdots \Delta q_n \rrangle$ featuring $3 \times n$ degrees of freedom. Measuring such features would then require collecting data in as many as $3 \times n$ dimensions. Clearly, a considerable reduction of this ``feature" space is required to enable the feasibility of measurements both in terms of data volumes (i.e., capturing sufficiently many $n$-tuples of particles to cover all partitions of the feature space with meaningful values) and in terms of its representation and  interpretation. While we do  not wish to preclude or dismiss possibilities of complex multi-dimensional analyses, we  focus the discussion, as a kind of extended motivation,
on some basic applications of the formalism and methods presented later in this work towards some specific physics analyses.

On general grounds, one can classify analyses of potential interest based on the number of kinematic partitions being used (i.e., partition of the $3\times n$ momentum space), the number of observables of interest (e.g., transverse momentum, $p_{\rm T}$, rapidity, $y$, charge, anisotropic flow coefficients, etc) and the number of particle types or species being considered (e.g., inclusive charged particles, positively vs. negatively charged particles, specific species such as pions, kaons, etc, and so on).  We thus organize the discussion in terms of few use cases, beginning with the simplest case involving a single variable $q$, and next considering progressively more and more complex use cases involving two variables: $q$, $p$, as well as several variables.

Analyses based on a single variable $q$ are already quite popular and have featured studies of fluctuations of transverse momentum, net-charge, etc. However, most prior analyses have been limited to two particles~\cite{Sharma:2008qr,STAR:2011iio,ALICE:2019smr,Sahoo:2022jgm} and only few recent works have undertaken higher number of particles~\cite{Voloshin:2002ku,ALICE:2023tej}. Notable exceptions to this statement  evidently include measurements of anisotropic flow based on multi-particle cumulants~\cite{Bilandzic:2010jr}.

Measurements of $p_{\rm T}$ (alternatively $y$ or $q$, etc) correlations
$\llangle \Delta p_{{\rm T},1} \Delta  p_{{\rm T},2} \cdots \Delta  p_{{\rm T},n} \rrangle$ involving $n \ge4$ particles of a given type of particle in a specific acceptance can be readily undertaken based on the methods discussed in sec.~\ref{sec:multiparticle}.
Various types of scaling are evidently possible to obtain dimensionless observables and assess the evolution of $n$-order momentum  correlators as a function of the  A--A collision centrality or produced particle multiplicity. An obvious choice is the inclusive momentum average $\llangle p_{\rm T}\rrangle$ already used in several studies~\cite{Sharma:2008qr,STAR:2011iio,ALICE:2019smr} but other choices of scaling have also been discussed~\cite{Giacalone:2020lbm}. Differential measurements can then be achieved, for instance,  by studying the magnitude of $\llangle \Delta p_{{\rm T},1} \Delta  p_{{\rm T},2} \cdots \Delta  p_{{\rm T},n} \rrangle$ vs. the width of the acceptance in rapidity.

Analyses involving two variables, $q$ and $p$, are of interest, for instance, towards the study of some specific observable (e.g.,  anisotropic flow, transverse momentum fluctuations, or net-charge fluctuations) in two kinematic partitions separated by a finite size rapidity gap.

\begin{figure}[ht]
  \centering
  \includegraphics[scale=0.35,keepaspectratio=true,clip=true,trim=2pt 2pt 2pt 2pt]
  {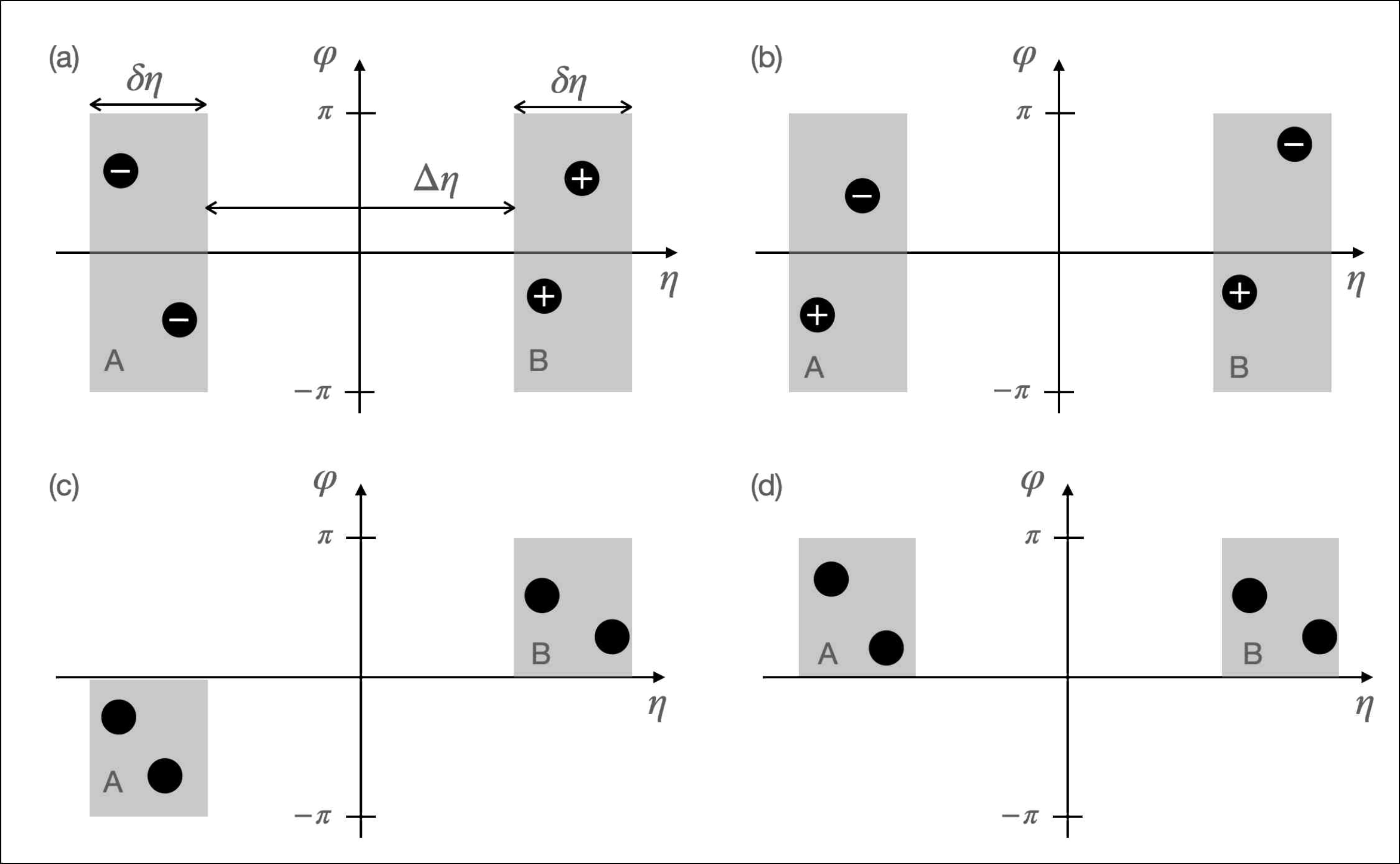} 
  \caption{\label{fig:rapidity-gap-geometry} Examples of acceptance configurations and particle selections are discussed in the text.}
\end{figure}

Considering examples of $p_{\rm T}$ fluctuation studies,  let $q_i$ and $p_i$ represent the transverse momentum of particles measured in two distinct rapidity acceptance ranges $\Omega_A$ and $\Omega_B$ of equal widths separated by a finite  rapidity gap $\Delta \eta$, as schematically illustrated in Fig.~\ref{fig:rapidity-gap-geometry}. One can then measure correlators of the form $\llangle \Delta q_1 \Delta p_1 \rrangle$,
$\llangle \Delta q_1 \Delta q_2 \Delta p_1\Delta p_2 \rrangle$, etc., at any  order to determine the strength of $n=2, 4$, etc,  transverse momentum correlations as a function of the width of the rapidity gap. Measurements of transverse momentum fluctuations and correlations have been thus far mostly limited to two-particle studies and implemented with a single acceptance bin~\cite{Broniowski:2005ae,CERES:2004bvl,Westfall:2004xy,NA49:2015lkz} or in a fully differential manner~\cite{STAR:2011iio,ALICE:2019smr}. The methods discused in Sec.~\ref{sec:multiparticle}, however, enable differential measurements involving multiple $n\ge 4$ particles. It then becomes possible to study momentum correlations arising from initial state fluctuations~\cite{Giacalone:2020lbm} while suppressing the influence of short range correlations (aka non-flow) associated with hadron decays and jet fragmentation. Letting $q_i$, $i=1, \ldots$, represent the $p_{\rm T}$ of particles in partition A, and $p_i$, $i=1, \ldots$ represent the $p_{\rm T}$ of particles in partition B, the described methods allow to obtain $\llangle q_1 q_2 \cdots q_n p_1 p_2 \cdots p_n \rrangle$ and the $n$-order deviates $\llangle \Delta q_1 \Delta q_2 \cdots \Delta q_n \Delta p_1 \Delta  p_2 \cdots \Delta p_n \rrangle$ corresponding to $p_{\rm T}$ correlators involving $n$ particles from partition A and $n$ particles from partition B. As in flow studies, one expects that it becomes possible to progressively suppress resonance and jet contributions by increasing the rapidity gap $\Delta\eta$ between partitions A an B. The analysis can also be made more differential by also using bins in azimuth as illustrated in panels (c) and (d) of Fig.~\ref{fig:rapidity-gap-geometry} thereby enabling the suppression of or focus on back-to-back jet contributions.

The methodology can readily be adopted also for measurements of charge correlations and, as it will
be introduced, multi-particle balance functions. In this case, $q_i$ and $p_i$ represent the charge of particles in partitions A and B. Then generic charge correlators $\llangle q_1 q_2 \cdots q_n p_1 p_2 \cdots p_n \rrangle$ and their deviates $\llangle \Delta q_1 \Delta q_2 \cdots \Delta q_n \Delta p_1 \Delta  p_2 \cdots \Delta p_n \rrangle$ can be obtaind and, it will be seen, these can then be related to multi-particle balance functions.

Two other  use cases based on two variables $q_i$ and $p_i$ are worth mentioning.
One involves the study of two distinct physics observables (e.g., charge, $p_{\rm T}$, rapidity, etc) in a single kinematic partition whereas the other involves the measurement of a specific particle observable, e.g., the $p_{\rm T}$, for two types of particle species. In the first case, the variable $q_i$ and $p_i$ represent the two observables of interest whereas in the second they tag the species of interest. These latter use cases  enable multi-particle correlations with specific species or between identical particles in two distinct $p_{\rm T}$ ranges.

The examples discussed in the previous paragraphs are readily extended towards the computation of correlation functions involving three or more kinematic partitions and particle types. Of particular interest is the determination of multiple particle balance functions. Although it may not be  practical to conduct analyses involving explicit computation of more than three or four kinematic partitions or species,  it  remains possible to consider balance functions involving large number of particles towards the study of long range multi-particle correlations constrained by charge conservation (or other quantum number conservation laws).
Let us  first  consider measurements of  four-particle balance functions based on  two kinematic partitions A and B separated by a finite rapidity gap, as illustrated in panels (a) and (b) of Fig.~\ref{fig:rapidity-gap-geometry}.
The partitions A and B could be azimuthally symmetric (i.e., with full azimuth coverage $0\le \varphi < 2\pi$), or feature partial coverage to suppress contributions from back-to-back jets, as schematically illustrated in panels (c) and (d) of the same figure.
Panel (a) illustrates a measurement involving two positively charged particles in A and two negatively particles in B. A measurement of the multi-particle balance function shall then be sensitive to the strength (or probability) of processes featuring four correlated particles separated by a finite rapidity gap.
Since 4-prong resonance decays are relatively few, this would reveal the likelihood of long range correlations determined by string-like fragmentation processes.
In contrast, the analysis illustrated in panel (b) would focus on correlated quartets featuring two nearby pairs of unlike sign particles. These could be produced by string-like fragmentation processes yielding four or more correlated particles, but they could also result from string fragmentation producing two neutral objects, each decaying into pairs of  $+$ve and $-$ve particles.  An explicitly selection of the charge states to be measured in partitions A and B might thus enable a discriminating study of the relative yields of  distinct processes. Indeed, an analysis of the dependence of the relative strengths of processes depicted in panel (a) and (b) could shed additional light on particle production process in elementary collisions. An analysis of the correlation strength performed as a function of the rapidity gap might then provide better sensitivity to the correlation length of string break up processes. Additionally, such  analyses conducted as a function of collision centrality and beam energy in large systems (A--A), and  comparisons to dependencies observed in small systems (e.g., pp and p--A), might then reveal whether this correlation length evolves with energy density, system size, collision energy, etc. Clearly, the position and size of measurement bins can be varied. Panel (c) illustrates a measurement geometry emphasizing back-to-back jet emission whereas panel (d) suppresses such processes and thus enables the study of long range non-jet and not resonance decay processes such as longitudinal string fragmentation. Obviously, a wide variety of other detection geometries can be implemented to explicitly favor or inhibit specific particle processes.

\begin{figure*}[htb]
\centering
\includegraphics[width=0.44\textwidth]{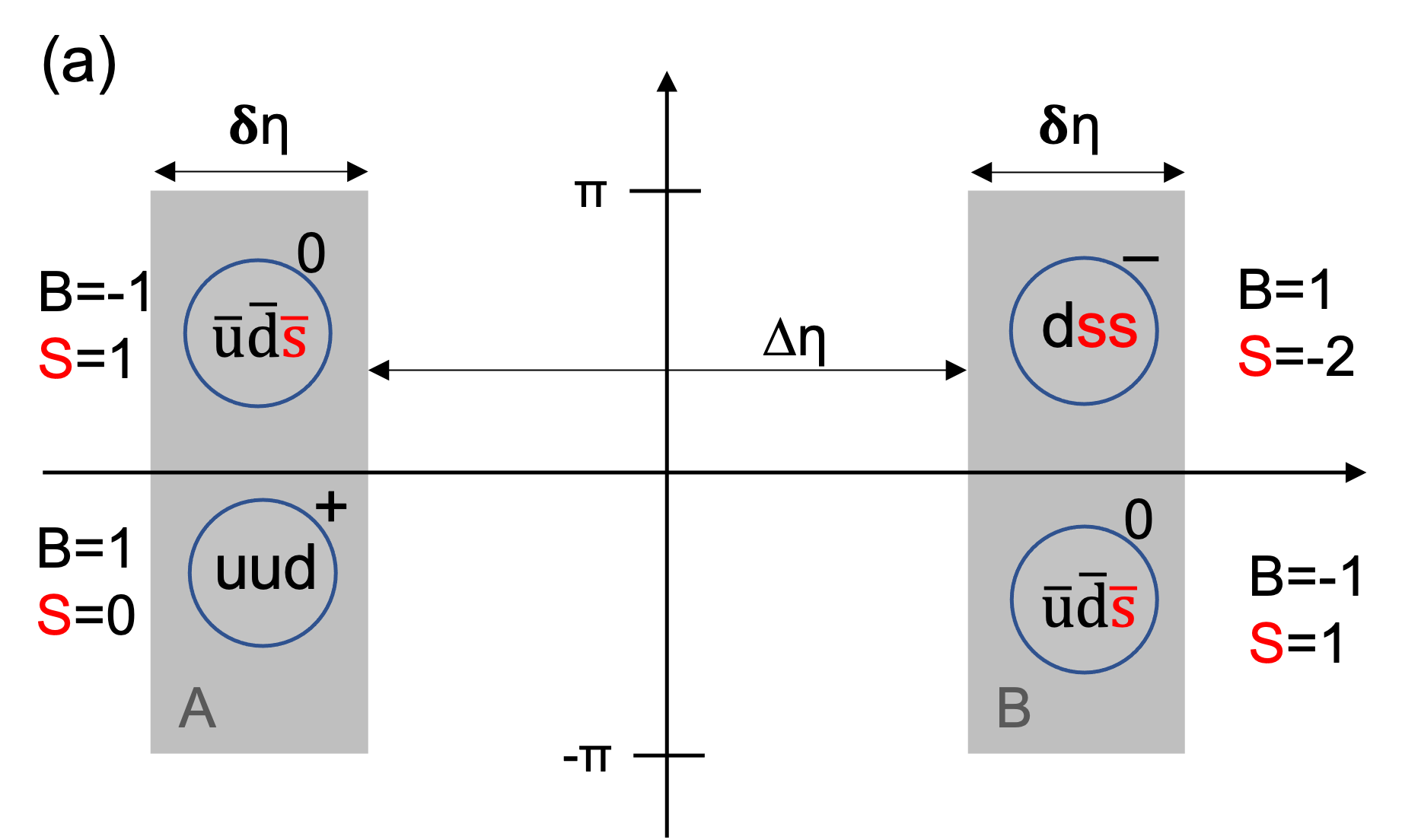}
~
\includegraphics[width=0.44\textwidth]{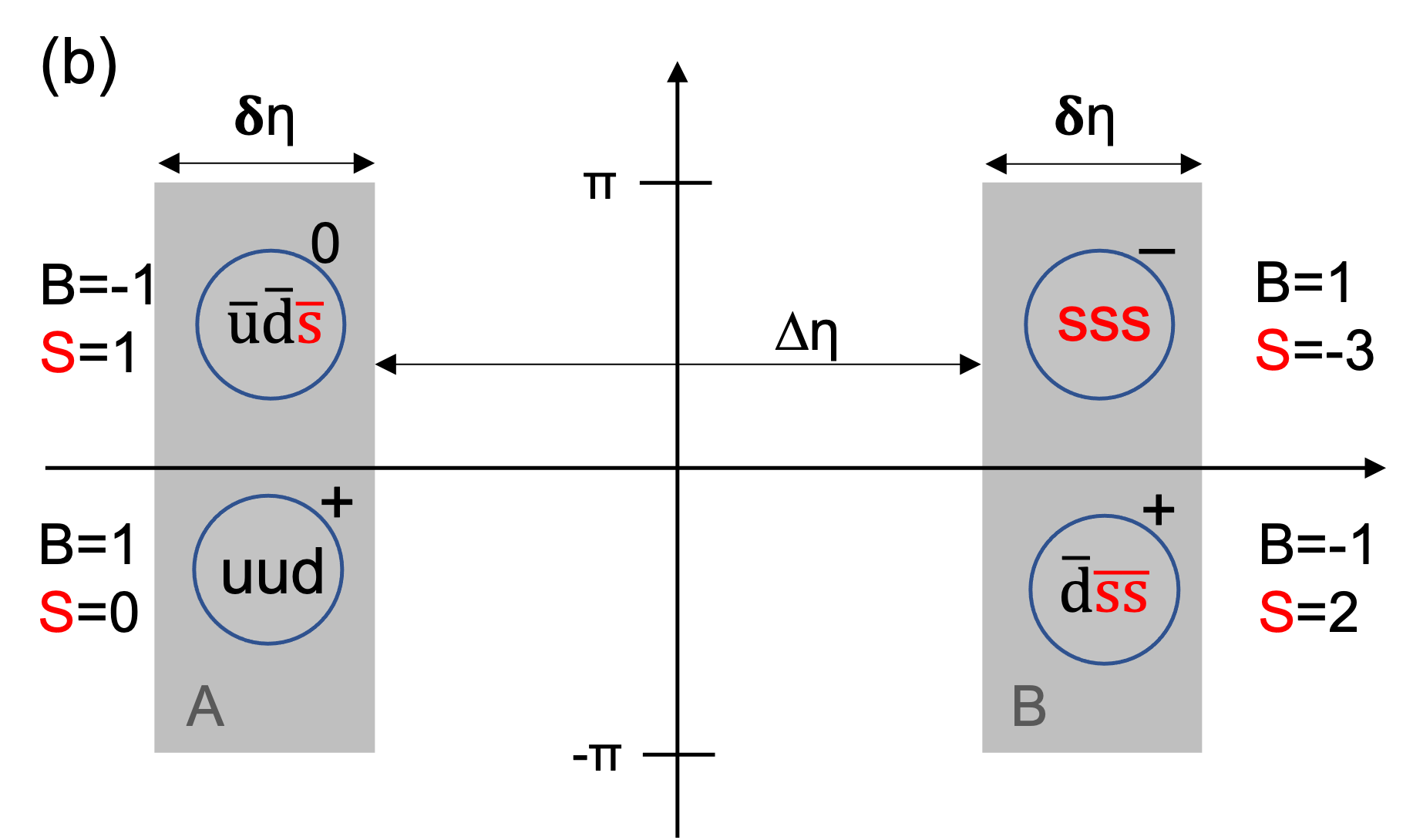}
\caption{Examples of multi-particles correlation for charge, baryon and strange balance studied using a rapidity gap.}
\label{fig:mpbf}
\end{figure*}

Analyses probing correlations of   three or four particles  of different charge, strangeness, and baryon number are also of interest and are possible with the framework presented in this paper.  For illustrative purposes, consider the  two scenarios displayed in Fig.~\ref{fig:mpbf}. Panel (a) illustrates a measurement involving a $\Xi (\rm dss)$ baryon and  a $\overline{\Lambda} (\overline{\rm uds})$ anti-baryon in rapidity  partition B, observed  jointly with a $\overline{\Lambda} (\overline{\rm uds})$ anti-baryon and a proton ($\rm uud$) in rapidity partition A, which simultaneously probes baryon and strangeness balancing. Similarly,  panel (b) shows a measurement involving a $\overline{\Lambda} (\overline{\rm uds})$ baryon and an proton ($\rm uud$) in partition A measured jointly with an 
$\Omega({\rm sss})$ baryon and a $\Xi$ baryon in partition B, which probes charge, strangeness, and baryon number balancing all at once.

In general, a large number of such particle species combinations could be combined to study   charge (Q), strangeness (S), baryon number (B), and isospin $I_{3}$ balancing. The tools could also be applied to  charmness (C) and/or bottomness (B) balancing. 

\section{Multi-particle Correlation Functions}
\label{sec:multiparticle}
\subsection{Cumulants and factorial cumulants}
\label{sec:cumulants}
As  already mentioned, in order to focus a study of correlated particles exclusively, one relies on the notions of integral and differential correlation function cumulants. Differential cumulants have been discussed in details elsewhere~\cite{Bilandzic:2010jr,Bilandzic:2013kga}. In the context of this work, it suffices to remember they can be ``reversed engineered" by listing all cluster decompositions of $n$-tuple densities. As such, single-, two-, and three-particle mixed  density cumulants  may be obtained by writing ~\cite{Pruneau:2017ypa}
\begin{align}
\label{eq:DiffCumulant1}
C_1^{\alpha}(\vec p_1) \equiv & \rho_1^{\alpha}(\vec p_1) \\
\label{eq:DiffCumulant2}
C_2^{\alpha_1\alpha_2}(\vec p_1,\vec p_2) \equiv &  \rho_2^{\alpha_1\alpha_2}(\vec p_1,\vec p_2) - 
C_1^{\alpha_1}(\vec p_1)C_1^{\alpha_2}(\vec p_2) =  \rho_2^{\alpha_1\alpha_2}(\vec p_1,\vec p_2) - 
\rho_1^{\alpha_1}(\vec p_1)\rho_1^{\alpha_2}(\vec p_2) \\ 
\label{eq:DiffCumulant3}
C_3^{\alpha_1\alpha_2\alpha_3}(\vec p_1,\vec p_2,\vec p_3) \equiv  &
\rho_3^{\alpha_1\alpha_2\alpha_3}(\vec p_1,\vec p_2,\vec p_3) - \sum_{\rm (3)} C_2^{\alpha_1\alpha_2}(\vec p_1,\vec p_2)C_1^{\alpha_3}(\vec p_3) -C_1^{\alpha_1}(\vec p_1)C_1^{\alpha_2}(\vec p_2)C_1^{\alpha_3}(\vec p_3),
\end{align}
where the notation $\sum_{\rm (k)}$ stands for a sum over $k$ ordered permutations of the labels $\alpha_1$, $\alpha_2$, and $\alpha_3$, as well as the corresponding momentum vectors $\vec p_1$, $\vec p_2$, and $\vec p_3$.  For $n=3$, substitution of first and second order cumulants and expansion of the permutations yield  
\begin{align}
\label{eq:DiffCumulant3V2}
C_3^{\alpha_1\alpha_2\alpha_3}(\vec p_1,\vec p_2,\vec p_3) 
=  &\rho_3^{\alpha_1\alpha_2\alpha_3}(\vec p_1,\vec p_2,\vec p_3) 
- \rho_2^{\alpha_1\alpha_2}(\vec p_1,\vec p_2)\rho_1^{\alpha_3}(\vec p_3)  
- \rho_2^{\alpha_1\alpha_3}(\vec p_1,\vec p_3)\rho_1^{\alpha_2}(\vec p_2)  \\ \nonumber
&- \rho_2^{\alpha_2\alpha_3}(\vec p_2,\vec p_3)\rho_1^{\alpha_1}(\vec p_1)  
+2\rho_1^{\alpha_1}(\vec p_1)\rho_1^{\alpha_2}(\vec p_2)\rho_1^{\alpha_3}(\vec p_3).
\end{align}
Higher order cumulants are computed and listed in Appendix~\ref{sec:MomentsCumulants}. 

Mixed cumulants $C_n^{\alpha_1\cdots\alpha_n}(\vec p_1,\,\ldots\,,\vec p_n)$ nominally feature $3\times n$ degrees of freedom and are, as such, challenging to measure and visually represent. The dimensionality,
and thus the number of degrees of freedom, can fortunately be reduced by integrating over several coordinates. When all coordinates are integrated within acceptances $\Omega_k$, $k=1,2,\ldots, n$, one obtains mixed factorial moments. For instance, the integration of $\rho_1^{\alpha}(\vec p)$ over a specific kinematic range  $\Omega$  yields the average  number of  particles of type $\alpha$ in this acceptance, whereas integration of two-, three-, or $n$ mixed particle densities yield the average number of pairs, triplets, and more generally $n$-tuplets of such groupings of particles. These averages are known as  mixed factorial moments, herewith denoted  $f_n^{\alpha_1\cdots \alpha_n}$, and computed according to
\begin{align}
\label{eq:f1}
f_1^{\alpha} \equiv& 
\int_{\Omega}\rho_1^{\alpha}(\vec p) {\rm d}\vec p = \langle N^{\alpha}\rangle, \\
\label{eq:f2}
f_2^{\alpha_1\alpha_2} \equiv&
\int_{\Omega_{\alpha_1}}\int_{\Omega_{\alpha_2}}\rho_2^{\alpha_1\alpha_2}(\vec p_1,\vec p_2) {\rm d}\vec p_1{\rm d}\vec p_2 
= \langle N^{\alpha_1}(N^{\alpha_2}-\delta_{\alpha_1\alpha_2})\rangle, \\
\label{eq:f3}
f_3^{\alpha_1\alpha_2\alpha_3} \equiv& \int_{\Omega_{\alpha_1}}\cdots\int_{\Omega_{\alpha_3}}
\rho_3^{\alpha_1\alpha_2\alpha_3}(\vec p_1,\vec p_2,\vec p_3) {\rm d}\vec p_1{\rm d}\vec p_2 {\rm d}\vec p_3  \\ \nonumber
=&\,\,\langle N^{\alpha_1}(N^{\alpha_2}-\delta_{\alpha_2\alpha_1})(N^{\alpha_3}-\delta_{\alpha_3\alpha_1}-\delta_{\alpha_3\alpha_2})\rangle, \\ 
f_n^{\alpha_1\,\cdots\,\alpha_n}\equiv& \int_{\Omega_{\alpha_1}}\cdots\int_{\Omega_{\alpha_3}}
\rho_n^{\alpha_1\cdots\alpha_n}(\vec p_1,\ldots,\vec p_n) {\rm d}\vec p_1\cdots {\rm d}\vec p_n \\ \nonumber
=&\,\,\left\langle 
N^{\alpha_1}
(N^{\alpha_2}-\delta_{\alpha_2\alpha_1})
\, \cdots\,
(N^{\alpha_n}-\delta_{\alpha_n\alpha_1}-\delta_{\alpha_n\alpha_2}- \cdots -\delta_{\alpha_n\alpha_{n-1}})
\right\rangle, 
\end{align}
where $\langle N^{\alpha}\rangle$, $\langle N^{\alpha_1}(N^{\alpha_2}-\delta_{\alpha_1\alpha_2}) \rangle$, and so on, denote the ensemble average of the number of mixed tuplets of the corresponding order.  Being integrals of densities, factorial moments do not discriminate correlated from uncorrelated particles and it is thus also convenient to consider mixed factorial moment cumulants, herein denoted $F_n^{\alpha_1\,\cdots\,\alpha_n}$. 
Mixed factorial moment cumulants, hereafter simply called factorial cumulants, are readily expressed in terms of integrals of the mixed cumulants $C_n^{\alpha_1\cdots\alpha_n}(\vec p_1,\ldots,\vec p_n)$ over acceptances $\Omega_k$, $k=1,\ldots, n$,  but can also be computed based on generating functions, as discussed in Appendix~\ref{sec:MomentsCumulants}. Lowest orders yield
\begin{align}
\label{eq:FactorialCumulant1}
F_1^{\alpha}=& \int C_1^{\alpha}(\vec p_1){\rm d}\vec p_1 = \langle N^{\alpha}\rangle = f_1^{\alpha}\\ 
\label{eq:FactorialCumulant2}
F_2^{\alpha_1\alpha_2}=& \int C_2^{\alpha_1\alpha_2}(\vec p_1,\vec p_2){\rm d}\vec p_1{\rm d}\vec p_2= f_2^{\alpha_1\alpha_2} -f_1^{\alpha_1}f_1^{\alpha_2} \\ 
\label{eq:FactorialCumulant3}
 F_3^{\alpha_1\alpha_2\alpha_3}=& f_3^{\alpha_1\alpha_2\alpha_3} - f_2^{\alpha_1\alpha_2}f_1^{\alpha_3} - f_2^{\alpha_1\alpha_3}f_1^{\alpha_2}
 - f_2^{\alpha_2\alpha_3}f_1^{\alpha_1} +2 f_1^{\alpha_1}f_1^{\alpha_2}f_1^{\alpha_3},
 \end{align}
while formula for higher orders are also listed in Appendix~\ref{sec:MomentsCumulants}.

\subsection{Joint Moments of Observables and their Deviates}
\label{sec:JointMoments}
As we discussed in Secs.~\ref{sec:motivation} and~\ref{sec:cumulants}, a wide variety of multi-particle correlation functions can be formulated in terms of the expectation value of products of particle observables of  the form $\llangle q_1 q_2 \, \ldots\, q_n \rrangle$ or their deviates $\llangle \Delta q_1 \Delta q_2 \, \ldots\,\Delta q_n \rrangle$.
In this section, we first introduce such expectations values based on moments of the sum $\sum_{i=0}^N q_i$ computed for all selected particles of an event with all self-correlations removed.  We then consider the expectation value of off-diagonal products of deviates of the form $\llangle \Delta q_1 \Delta q_2 \, \cdots\, \Delta  q_n\rrangle$. First note that these expressions are totally general and thus applicable for any type of particle observables, e.g., transverse momentum, rapidity, electric charge, or other quantum numbers. Together with formula introduced in sec.~\ref{sec:method} and Appendix~\ref{sec:DeviatesAverage}, these correlators enable the formulation of both integral and differential measurements of multiple particle correlations of basic particle observables.

In order to obtain expressions sought for, first consider a  particle observable of interest $q$. This   observable could be the transverse momentum of the particle, its rapidity, its charge, some other quantum numbers, or simply unity (i.e., to count the particles). We will  denote the value of this observable for a specific particle    $q_i$, with the index $i$ spanning all selected particles (i.e., satisfying  specific kinematic and quality selection criteria)  in a given event.
We are interested in computing moments  of $q$ and its deviates $\Delta q_i \equiv q_i - \llangle q\rrangle$, where $\llangle q\rrangle$ is the inclusive event ensemble average of $q$ for specific collision conditions (i.e., events satisfying specific selection criteria).

Inclusive event ensemble (joint) averages of products of $q$s are defined according to
\begin{align}
\label{eq:inclusiveAvg-q}
\llangle  q \rrangle =& \frac{1}{\langle N\rangle}\left\langle \sum_{i=1}^N q_i \right \rangle, \\
\label{eq:inclusiveAvg-q1q2}
\llangle  q_{1} q_{2}\rrangle =& \frac{1}{\langle N(N-1)\rangle} \left\langle \sum_{i_1\ne i_2=1}^N q_{i_1} q_{i_2}\right\rangle, \\
\label{eq:inclusiveAvg-q1q1...qn}
\llangle  q_{1} q_{2} \cdots  q_{n}\rrangle =&
\frac{1}{\langle N(N-1)\,\cdots\, (N-n+1)\rangle}  \left\langle  \sum_{i_1\ne i_2\ne \cdots \ne i_n=1}^N q_{i_1} q_{i_2} \cdots  q_{i_n}\right\rangle,
\end{align}
where the sums run over all $N$  selected particles in a given event. The notation $\sum_{i_1\ne i_2}$ indicates the sums  are computed for distinct  particles, i.e., distinct values of $i_1$, $i_2$, etc. As such,  $\sum_{i=1}^N q_i$ represents the sum of the $q_i$ in a given event, and $\langle \sum_{i=1}^N q_i \rangle$ is the event ensemble average of this sum across all events of a selected data sample.  Similarly,
$\langle \sum_{i\ne j=1}^N q_i q_j\rangle$, and higher orders, represent ensemble averages of products of the $q$s evaluated for $n$-tuplets of  particles. However, the fact that sums proceeds on $i\ne j$ implies auto-correlations are explicitly removed. Herewith, we will use inclusive averages (i.e., averages computed over an event ensemble) but it is trivial to change the definition to event-by-event averages~\cite{Giacalone:2020lbm}. See for instance Eqs.~(\ref{eq:Dq1Dq2VsQeventwise} - \ref{eq:Dq1Dq2Dq3VsQeventwise}).
Additionally, note that if it is possible to pre-scan the dataset of interest to determine the inclusive average $\llangle q\rrangle$, then one can readily replace $q_i$ by $\Delta q_i=q_i-\llangle q\rrangle$ in Eqs.~(\ref{eq:inclusiveAvg-q}--\ref{eq:inclusiveAvg-q1q1...qn}) instead of carrying out the factorization discussed in details below.

We proceed to compute event ensemble averages of moments of $\Delta q_i$ in terms of moments of $q_i$.  The inclusive first moment of $\Delta q$ is calculated according to
\begin{align}
\label{eq:inclusive avg of Dq}
    \llangle \Delta q \rrangle =& \frac{1}{\langle N \rangle} \left \langle \sum_{i=1}^N \Delta q_i  \right\rangle = \frac{1}{\langle N \rangle}  \sum_{i=1}^N  q_i  - \frac{\langle N \rangle}{\langle N \rangle} \llangle q \rrangle = \llangle q \rrangle  - \llangle q \rrangle  =0
\end{align}
and vanishes by construction. In order to compute moments of order $n=2$ and $n=3$, we write 
\begin{align}
\label{eq:deviate2}
    \Delta q_i \Delta q_j =& q_i q_j - \llangle  q \rrangle (q_i + q_j) + \llangle q \rrangle^2, \\
\label{eq:deviate3}
    \Delta q_i \Delta q_j \Delta q_k =& q_i q_j q_k - \llangle  q \rrangle \left(q_iq_j + q_iq_k + q_jq_k \right)
    +\llangle  q \rrangle^2 \left(q_i+q_j+q_k \right) - \llangle q \rrangle^3.
\end{align}
whereas higher order products can be written
\begin{align}
    \label{eq:deviateOfN}
    \Delta q_{i_1} \Delta q_{i_2}\, \cdots\, \Delta q_{i_n}=&\, q_{i_1} q_{i_2} \,\cdots\, q_{i_n} - \llangle  q \rrangle \sum_{\binom{n}{1}} q_{i_1} q_{i_2} \cdots q_{i_{n-1}}
    +\llangle  q \rrangle^2 \sum_{\binom{n}{2}} q_{i_1} q_{i_2} \cdots q_{i_{n-2}}  \\ \nonumber
    &-\llangle  q \rrangle^3 \sum_{\binom{n}{3}} q_{i_1} q_{i_2} \cdots q_{i_{n-3}}  \,\,
    \cdots \,\, + (-1)^{n-1}\llangle q \rrangle^{n-1}\sum_{\binom{n}{n-1}} q_{i_1}
    +(-)^n \llangle q \rrangle^n,
\end{align}
where the notation $\sum_{\binom{n}{k}}$ indicates a sum over all $\binom{n}{k}$  permutations of the indices $i_1$, $i_2$, etc.
Ensemble averages of products of order $n=2,\, 3,\, 4$ yield
\begin{align}
\label{eq:inclusive avg of Dq_i Dq_j}
 \llangle \Delta q_i \Delta q_j\rrangle =&  \frac{1}{\langle N(N-1)\rangle} \left\langle \sum_{i\ne j=1}^N \left( q_i q_j - \llangle q \rrangle (q_i + q_j) + \llangle  q \rrangle^2 \right)  \right\rangle =   \llangle  q_i q_j\rrangle -  \llangle  q \rrangle^2, \\
\label{eq:inclusive avg of Dq_i Dq_j Dq_k}
 \llangle  \Delta q_i \Delta q_j\Delta q_k\rrangle =&  \llangle  q_i q_jq_k\rrangle - 3 \llangle  q \rrangle \llangle  q_i q_j \rrangle + 2\llangle  q \rrangle^3, \\
 \label{eq:inclusive avg of Dq_i Dq_j Dq_k Dq_l}
 \llangle     \Delta q_i \Delta q_j\Delta q_k\Delta q_l\rrangle =&  \llangle  q_i q_jq_kq_l\rrangle
 - 4 \llangle  q \rrangle \llangle  q_i q_j q_k\rrangle
 + 6 \llangle  q \rrangle^2 \llangle  q_i q_j\rrangle - 3\llangle  q \rrangle^4,
\end{align}
whereas higher orders can be computed according to
\begin{align}
\label{eq:inclusive avg of DqOfN}
 \llangle     \Delta q_{i_1} \Delta q_{i_2}\,\cdots\, \Delta q_{i_n}\rrangle =&  \llangle  q_{i_1} q_{i_2}\,\cdots\, q_{i_n} \rrangle
- \binom{n}{n-1} \llangle  q \rrangle \llangle q_{i_1} q_{i_2} \cdots q_{i_{n-1}}\rrangle
 + \binom{n}{n-2} \llangle  q \rrangle^2   \llangle q_{i_1} q_{i_2} \cdots q_{i_{n-2}}\rrangle \\ \nonumber
 &+\,\cdots\, (-1)^{n-2}\binom{n}{2} \llangle  q \rrangle^{n-2} \llangle  q_{i_1} q_{i_2} \rrangle
 +(-1)^{n-1} \binom{n}{1}\llangle  q \rrangle^{n-1} \llangle  q_1 \rrangle +(-1)^{n} \binom{n}{0}\llangle  q \rrangle^n \\
 =& \sum_{k=0}^n  (-1)^{n-k} \binom{n}{k} \llangle q\rrangle^{n-k} \llangle  q_{i_1}\,\cdots\,q_{i_k} \rrangle,
\end{align}
where $\llangle  q_{i_1}\,\cdots\,q_{i_k} \rrangle=\llangle q\rrangle$ for $k=1$
and $\llangle  q_{i_1}\,\cdots\,q_{i_k} \rrangle=1$ for $k=0$.

The computation of moments of order $n$ of $q$s amounts to sums of the form $\sum_{i\ne j=1}^N q_i q_j$, $\sum_{i\ne j \ne k=1}^N q_i q_jq_k$, etc,
which nominally require up to $n$ nested loops for each event considered. Though conceptually trivial, such calculations involve a  computation time proportional to $N^n$ that  becomes quickly prohibitive for large values of the multiplicity $N$. Fortunately, the method of moments, which we discuss in the next two sub-sections, enables the computation of averages of these sums based on a single loop (per event), i.e., with a computation time proportional to $N$.

\subsection{Method of Moments}
\label{sec:method}

The method of moments is introduced to facilitate the computation of moments $\llangle q_1 q_2 \, \ldots\, q_n \rrangle$ and  deviates $\llangle \Delta q_1 \Delta q_2 \, \ldots\,\Delta q_n \rrangle$ based on a single loop over all particles of an event rather than using nested loops.

\subsubsection{Method of Moments for a single variable}
\label{sec:method-single}

Let $Q_n$ represent an event-wise sum of the $n$th power of the variable $q_i$ of the $N$ selected particles of an event
\begin{equation}
\label{eq:Qn}
Q_n = \sum_{i=1}^N q_i^n.
\end{equation}
The method of moments relies on the evaluation of ensemble averages of  $Q_n$  and products of the form
$Q_nQ_m$, $Q_nQ_mQ_o$, etc.
The (inclusive) ensemble average of $Q_n$ is
\begin{equation}
\label{eq:avg Qn}
  \llangle  Q_n \rrangle= \frac{1}{\langle N \rangle} \llangle q^n\rrangle,
\end{equation}
and one  obviously obtains
\begin{equation}
\label{eq:avg qn}
  \llangle q^n\rrangle = \langle N \rangle\llangle  Q_n \rrangle.
\end{equation}
Computation of the ensemble average of products $Q_nQ_m$, $Q_nQ_mQ_o$, etc, requires one properly handles the expansions of the sums. For instance, product of two and three $Q$s yield
\begin{align}
\label{eq:Q_n Q_m}
Q_{n_1} Q_{n_2}  =& \sum_{i_1,i_2=1}^N q_{i_1}^{n_1} q_{i_2}^{n_2} = \sum_{i=1}^N q_i^{n_1+n_2} + \sum_{i_1\ne i_2=1}^N q_{i_1}^{n_1}q_{i_2}^{n_2}, \\
\label{eq:Q_n Q_m Q_o}
Q_{n_1} Q_{n_2} Q_{n_3} =& \sum_{i=1}^N q_i^{n_1+n_2+n_3}
+ \sum_{i_1\ne i_2=1}^N q_{i_1}^{n_1+n_2}q_{i_2}^{n_3}
+ \sum_{i_1\ne i_2=1}^N q_{i_1}^{n_1+n_3}q_{i_2}^{n_2}
+ \sum_{i_1\ne i_2=1}^N q_{i_1}^{n_1}q_{i_2}^{n_2+n_3} \\ \nonumber
&+ \sum_{i_1\ne i_2\ne i_3=1}^N q_{i_1}^{n_1} q_{i_2}^{n_2} q_{i_3}^{n_3}, \\
=& \sum_{i=1}^N q_i^{n_1+n_2+n_3} + \sum_{(3)} \sum_{i_1\ne i_2=1}^N q_{i_1}^{n_1+n_2}q_{i_2}^{n_3} + \sum_{i_1\ne i_2\ne i_3=1}^N q_{i_1}^{n_1} q_{i_2}^{n_2} q_{i_3}^{n_3},
\end{align}
where the notation $\sum_{(3)}$ represents a sum spanning all three ordered permutations of $n_1$, $n_2$, and $n_3$.

Clearly, calculation of the ensemble average of   $Q_{n_{1}}Q_{n_{2}}$ yields
\begin{align}
\label{eq:Q2avg}
\llangle Q_{n_{1}}Q_{n_{2}}\rrangle  =& \langle N\rangle \llangle q^{n_{1}+n_{2}}\rrangle  + \langle N\left(N-1\right)\rangle \llangle q_1^{n_{1}}q_2^{n_{2}}\rrangle.
\end{align}
This expression contains a term of the form $\langle N\rangle \llangle q_{i_{1}}^{n_{1}+n_{2}}\rrangle$ that can be readily replaced by $\llangle Q_{n_1+n_{2}}\rrangle$ based on Eq.~(\ref{eq:avg Qn}). Solving for $\llangle  q_1^{n_1} q_2^{n_2} \rrangle$, one then gets
\begin{align}
\label{eq:avg q_i^n q_j^m}
  \langle N(N-1) \rangle \llangle  q_1^{n_1} q_2^{n_2} \rrangle= \llangle Q_{n_1}Q_{n_2}\rrangle -  \llangle Q_{n_1+n_2}\rrangle.
\end{align}
which for $n_1=n_2=1$ evidently simplifies to
\begin{equation}
\label{eq:avg q_i q_j}
\langle N(N-1) \rangle  \llangle  q_1 q_2\rrangle
= \llangle Q_1^2\rrangle -  \llangle Q_{2}\rrangle.
\end{equation}
Proceeding similarly for the ensemble average of $Q_{n_{1}}Q_{n_{2}}Q_{n_{3}}$, one gets
\begin{align}
\label{eq:Q3avg}
\llangle Q_{n_{1}}Q_{n_{2}}Q_{n_{3}}\rrangle  =& \langle N\rangle \llangle q_{i_{1}}^{n_{1}+n_{2}+n_{3}}\rrangle
+ \langle N\left(N-1\right)\rangle \llangle q_{i_{1}}^{n_{1}+n_{2}}q_{i_{2}}^{n_{3}}\rrangle
+ \langle N\left(N-1\right)\rangle \llangle q_{i_{1}}^{n_{1}+n_{3}}q_{i_{2}}^{n_{2}}\rrangle \\ \nonumber
&+ \langle N\left(N-1\right)\rangle \llangle q_{i_{1}}^{n_{2}+n_{3}}q_{i_{2}}^{n_{1}}\rrangle
+ \langle N\left(N-1\right)\left(N-2\right)\rangle \llangle q_{i_{1}}^{n_{1}}q_{i_{2}}^{n_{2}}q_{i_{3}}^{n_{3}}\rrangle.
\end{align}
The first term, $\langle N\rangle \llangle q_{i_{1}}^{n_{1}+n_{2}+n_{3}}\rrangle$, is equal to $\llangle Q_{n_1+n_2+n_3}\rrangle$, while the next three terms are of the form of Eq.~(\ref{eq:avg q_i^n q_j^m}). Substituting these terms and solving for  $\llangle q_{i_{1}}^{n_{1}}q_{i_{2}}^{n_{2}}q_{i_{3}}^{n_{3}}\rrangle$, one gets
\begin{align}
\label{eq:avg q_i^n q_j^m q_k^o}
  \langle N(N-1)(N-2)\rangle \llangle  q_1^{n_{1}} q_1^{n_{2}}q_1^{n_{3}} \rrangle =&
\llangle Q_{n_{1}}Q_{n_{2}}Q_{n_{3}}\rrangle
- \llangle Q_{n_{1}+n_{2}}Q_{n_{3}}\rrangle
- \llangle Q_{n_{1}+n_{3}}Q_{n_{2}}\rrangle
- \llangle Q_{n_{2}+n_{3}}Q_{n_{1}}\rrangle  \\ \nonumber
&+2 \llangle Q_{n_{1}+n_{2}+n_{3}}\rrangle
\end{align}
which, for $n_1=n_2=n_3=1$, reduces to
\begin{align}
 \langle N(N-1)(N-2)\rangle
\llangle  q_1 q_2 q_3 \rrangle =& \llangle  Q_1^3  \rrangle -3 \llangle  Q_{2}Q_1 \rrangle  + 2 \llangle  Q_{3} \rrangle,
\end{align}
The above calculation can be repeated iteratively for higher order products of $q$s and  thus yield expressions for
 $\llangle q_1  q_2 \,\cdots \, q_n \rrangle$ at arbitrarily high order $n$. In practice, such calculations become rather tedious for  $n>4$ and are best computed programmatically, as discussed in Appendix~\ref{sec:computation}.

\subsubsection{Higher Order Moments of  Mixed Acceptance Variates }
\label{sec:method-mixed}

In the previous sections, we considered calculations of  higher moments  $\llangle q_1 q_2 \cdots q_n \rrangle$ computed for  a single acceptance or particle species (i.e., for a single kinematic bin or for a specific species or both). In order to compute differential correlation functions involving several kinematic bins or species, we now proceed to compute expectation values of the form $\llangle q_1 q_2 \cdots q_n p_1 \cdots p_m r_1 \cdots r_o\rrangle$ where $q$, $p$, and $r$ represent distinct kinematic bins or species.
The discussion is here limited to three bins (or species) for simplicity's sake but it is trivially extended to an arbitrary number of bins and species. The particle multiplicities in each bin are denoted $N_i$, $i=1,\ldots, 3$. Deviates are denoted $\Delta q_i$, $\Delta p_j$, and $\Delta r_k$  for particles in bins 1, 2, and 3, respectively. Moments for all particles detected  in a single bin are given by expressions of the form~(\ref{eq:avg q_i^n q_j^m} - \ref{eq:avg q_i^n q_j^m q_k^o}) already considered in sec.~\ref{sec:method-single}. We thus need to consider mixed moments only in this section. Lowest mixed moments are given by expressions of the form
\begin{align}
\label{eq:q(1)p(1)}
\llangle  q_1 p_1\rrangle =& \frac{1}{\langle N_1N_2\rangle } \left\langle \sum_{i=1}^{N_1} q_i \sum_{j=1}^{N_2}  p_j\right\rangle, \\
\label{eq:q(2)p(1)}
\llangle  q_1 q_2 p_1\rrangle =& \frac{1}{\langle N_1(N_1-1)N_2\rangle }  \left\langle \sum_{i_1\ne i_2=1}^{N_1} q_{i_1} q_{i_2} \sum_{j=1}^{N_2}  p_{j}\right\rangle \\
\label{eq:q(1)p(1)r(1)}
\llangle  q_1 p_1 r_1\rrangle =& \frac{1}{\langle N_1N_2N_3\rangle }  \left\langle  \sum_{i=1}^{N_1} q_i \sum_{j=1}^{N_2} p_j \sum_{k=1}^{N_3} r_k\right\rangle.
\end{align}
More generally, considering moments of order $m_1$, $m_2$, and $m_3$ for bins 1, 2, and  3, one gets expressions of the form
\begin{align}
\label{eq:q(m)p(m)r(m)}
\llangle  q_1\,\cdots\,q_{m_1} p_1\,\cdots\,p_{m_2}r_1\,\cdots\,r_{m_3}\rrangle =&
\frac{1}{N_{m_1,m_2,m_3}}
\left\langle
\sum_{i_1\ne \cdots\ne  i_{m_1}=1}^{N_1} q_{i_1}q_{i_2}\cdots\, q_{i_{m_1}}
\sum_{j_1\ne \cdots\ne  j_{m_2}=1}^{N_2} p_{j_1}p_{j_2}\cdots\, p_{j_{m_2}} \right. \\ \nonumber
& \left.
\times \sum_{k_1\ne \cdots\ne  k_{m_3}=1}^{N_3} r_{k_1}r_{k_2}\cdots\, r_{k_{m_3}}
\right\rangle,
\end{align}
where the sums $\sum_{i_1\ne \cdots\ne  i_{m_1}=1}^{N_1}$, $\sum_{j_1\ne \cdots\ne  j_{m_2}=1}^{N_2}$  and $\sum_{k_1\ne \cdots\ne  k_{m_3}=1}^{N_3}$ span tuplets of distinct particles  in bins 1, 2, and 3, respectively. Also note that  the normalization corresponds to the average number of such tuplets:
\begin{align}
\label{eq:norm}
N_{m_1,m_2,m_3} \equiv& \langle
N_1 (N_1-1) \cdots (N_1-m_1+1) N_2 (N_2-1) \cdots (N_2-m_2+1)  N_3 (N_3-1) \cdots (N_3-m_3+1) \rangle.
\end{align}

Event ensembles of mixed moments of deviates are defined in a similar fashion. At lowest orders, one gets
expressions of the form
\begin{align}
\label{eq: Delta q(1) Delta p(1)}
\llangle  \Delta q_1 \Delta  p_1\rrangle =&  \llangle q_1 p_1\rrangle - \llangle q\rrangle\llangle p\rrangle, \\
\label{eq: Delta q(1) Delta p(1) Delta r(1)}
\llangle \Delta  q_1 \Delta p_1 \Delta r_1\rrangle =&
\llangle  q_1 p_1 r_1\rrangle
-\llangle p\rrangle\llangle  q_1 r_1\rrangle
-\llangle q\rrangle\llangle  p_1 r_1\rrangle
-\llangle r\rrangle\llangle  q_1 p_1\rrangle 
+2\llangle q\rrangle\llangle  p\rrangle\llangle  r\rrangle,
\end{align}
and higher order moments are discussed in Appendix~\ref{sec:DeviatesAverage}.  The computation of mixed moments and their deviates nominally requires
nested loops over all particles and bins of interest. But as for the single bin case discussed in the previous section,  it is advantageous to introduce event-wise sums of the variables $q_i$, $p_i$, and $r_i$  according to
\begin{align}
    Q_n =& \sum_{i=1}^N q_i^n, \hspace{0.3in}
    P_n = \sum_{i=1}^N p_i^n, \hspace{0.3in}
    R_n = \sum_{i=1}^N r_i^n.
\end{align}
It then becomes possible, as discussed in Appendix~\ref{sec:DeviatesAverage}, to  compute the event ensemble averages $\llangle  q_1\,\cdots\,q_{m_1}$ $ p_1\,\cdots\,p_{m_2}r_1\,\cdots\,r_{m_3}\rrangle$ and their corresponding
deviates $\llangle  \Delta q_1\,\cdots\,\Delta q_{m_1} \Delta p_1\,\cdots\,\Delta p_{m_2}\Delta r_1\,\cdots\,\Delta r_{m_3}\rrangle$ based on recursive formula of the moments of $Q_n$, $P_n$, and $R_n$.

\subsection{Differential Measurements of Multiple-Particle Correlations}
\label{sec:configurations}

The multi-particle correlators $\llangle q_1 q_2 \cdots q_n \rrangle$ and
$\llangle \Delta q_1 \Delta  q_2 \cdots \Delta q_n \rrangle$  presented in sec.~\ref{sec:cumulants}, equipped with the method of moments discussed in sec.~\ref{sec:method},  provide a basis for the  extension of former techniques used for measurements of transverse momentum correlations, rapidity correlations, as well as charge  correlations (including baryon and strangeness numbers correlations).

The method of moments, whether used with a single or several variables, is nominally very powerful because it enables joint measurements involving many particles simultaneously. However, it is also clear that the complexity of such measurements can quickly grow out of hand.

As was described in Sec.~\ref{sec:motivation}, one can classify analyses of potential interest based on the number of kinematic bins being used (i.e., partitions of the $3\times$ momentum space), the number of observables of interest (e.g., transverse momentum, $p_{\rm T}$, rapidity, $y$, charge $q$, anisotropic flow coefficients, etc) and the number of particle types or species being considered (e.g., inclusive charged particles, positively vs. negatively charged particles, specific species such as pions, kaons, etc, and so on).  We thus organize the discussion of this section in terms of few use cases, beginning with the simplest case involving a single variable $q$, with event-wise variable $Q_n$, and next considering progressively more complex use cases involving two  variables: $q$, $p$ with event-wise variables $Q_n$ and $P_n$, as well as more complex analyses based on several variables.

\subsubsection{One variable (\texorpdfstring{$q$}\xspace)}

Measurements of $p_{\rm T}$ (alternatively $y$ or $q$, etc) correlations
$\llangle \Delta p_{{\rm T},1} \Delta  p_{{\rm T},2} \cdots \Delta  p_{{\rm T},n} \rrangle$ involving $n \ge4$ particles of a given type of particle in a specific acceptance can be readily undertaken based on the method of moments discussed in sec.~\ref{sec:method}. If it is possible or practical
to carry out the analysis in two or more passes on the data, than one can use the first pass to determine $\llangle p_{\rm T}\rrangle$. In the second pass, one can then
define and compute $Q_n = \sum_i^N (p_{\rm T}-\llangle p_{\rm T}\rrangle)$
event by event and then use  Eqs.~(\ref{eq:q1q2VsQs} - \ref{eq:q1q2q3q4q5q6q7q8VsQs}) to obtain $n$-oder moments $\llangle \Delta p_{{\rm T},1} \Delta  p_{{\rm T},2} \cdots \Delta  p_{{\rm T},n} \rrangle$. If the determination of $\llangle p_{\rm T}\rrangle$ in a first pass is not practical, then one can define and compute $Q_n = \sum_i^N p_{\rm T}$
event by event, use Eqs.~(\ref{eq:q1q2VsQs} - \ref{eq:q1q2q3q4q5q6q7q8VsQs}) to obtain $\llangle p_{{\rm T},1} p_{{\rm T},2} \cdots p_{{\rm T},n} \rrangle$
and Eqs.~(\ref{eq:Dq1Dq2} - \ref{eq:Dq1Dq2Dq3Dq4Dq5Dq6}) or Eq.~(\ref{eq:Dq1-Dqm}) to obtain the $n$-order deviates $\llangle \Delta p_{{\rm T},1} \Delta  p_{{\rm T},2} \cdots \Delta  p_{{\rm T},n} \rrangle$.

\subsubsection{Two variables (\texorpdfstring{$q$ and $p$}\xspace)}

We first discuss examples of $p_{\rm T}$ fluctuation studies.  Let $q_i$ and $p_i$ represent the transverse momentum of particles measured in two distinct rapidity acceptance ranges $\Omega_A$ and $\Omega_B$ of equal widths separated by a finite  rapidity gap $\Delta \eta$, as schematically illustrated in Fig.~\ref{fig:rapidity-gap-geometry}. One can then measure correlators of the form $\llangle \Delta q_1 \Delta p_1 \rrangle$,
$\llangle \Delta q_1 \Delta q_2 \Delta p_1\Delta p_2 \rrangle$, etc., at any  order to determine the strength of $n=2, 4$, etc,  transverse momentum correlations as a function of the width of the rapidity gap. The method of moments, however, enables differential measurements involving multiple $n\ge 4$ particles. Let $q_i$, $i=1, \ldots$, represent the $p_{\rm T}$ of particles in bin A, and $p_i$, $i=1, \ldots$ represent the $p_{\rm T}$ of particles in bin B. One then defines event-wise variables $Q_n = \sum_{i}^N p_{{\rm T},i}^n$ (from bin A) and
$P_n = \sum_{i}^N p_{{\rm T},i}^n$ (from bin B). Equations~(\ref{eq:q1q2VsQs} - \ref{eq:q1q2q3q4q5q6q7q8VsQs}) are then used to obtain $\llangle q_1 q_2 \cdots q_n p_1 p_2 \cdots p_n \rrangle$
and Eqs.~(\ref{eq:Dq1Dq2} - \ref{eq:Dq1Dq2Dq3Dq4Dq5Dq6}) or Eq.~(\ref{eq:Dq1-Dqm}) are used to obtain the $n$-order deviates $\llangle \Delta q_1 \Delta q_2 \cdots \Delta q_n \Delta p_1 \Delta  p_2 \cdots \Delta p_n \rrangle$ corresponding to $p_{\rm T}$ correlators involving $n$ particles from bin A and $n$ particles from bin B.

The method described above can readily be adopted also for measurements of charge correlations and multi-particle balance functions. In this case, $q_i$ and $p_i$ represent the charge of particles in bins A and B. Applications of Eqs.~(\ref{eq:q1q2VsQs} - \ref{eq:q1q2q3q4q5q6q7q8VsQs}) then yield generic charge correlators $\llangle q_1 q_2 \cdots q_n p_1 p_2 \cdots p_n \rrangle$
and Eqs.~(\ref{eq:Dq1Dq2} - \ref{eq:Dq1Dq2Dq3Dq4Dq5Dq6}) or Eq.~(\ref{eq:Dq1-Dqm}) can be used to compute deviates.

Two additional use cases based on two variables $q_i$ and $p_i$ (with the corresponding event-wise variables $Q_n$ and $P_n$)  are worth mentioning.
One involves the study of two distinct physics observables (e.g., charge, $p_{\rm T}$, rapidity, etc) in a single rapidity bin whereas the other involves the measurement of a specific particle observable, e.g., the $p_{\rm T}$, for two types of particle species. In the first case, the variable $q_i$ and $p_i$ represent the two observables of interest whereas in the second they tag the species of interest. The determination of correlators between two observables or two types of particle species then proceeds in the manner already described. First get  $\llangle q_1 q_2 \cdots q_n p_1 p_2 \cdots p_n \rrangle$ with Eqs.~(\ref{eq:q1q2VsQs} - \ref{eq:q1q2q3q4q5q6q7q8VsQs}) and next used Eqs.~(\ref{eq:Dq1Dq2} - \ref{eq:Dq1Dq2Dq3Dq4Dq5Dq6}) to compute the deviates of interest.

\subsubsection{Three or more  variables 
(\texorpdfstring{$q$}\xspace,
\texorpdfstring{$p$}\xspace,
\texorpdfstring{$r$}\xspace, ...)}

The examples discussed in the previous paragraph are readily extended towards the computation of factorial cumulants or correlation functions involving three or more kinematic bins and particle types. Of particular interest is the determination of multiple particle balance functions. Although it may not be  practical to conduct analyses involving explicit computation of more than three or four kinematic bins or species,  it  remains possible to consider balance functions involving large number of particles towards the study of long range multi-particle correlations constrained by charge conservation (or other quantum number conservation laws).

As mentioned in Sec.~\ref{sec:motivation}, we consider here the study of four-particle balance functions using two kinematic bins $A$ and $B$ separated by a finite rapidity gap, as was illustrated in panels (a) and (b) of Fig.~\ref{fig:rapidity-gap-geometry}.
The bins  $A$ and $B$ could be azimuthally symmetric (i.e., with full azimuth coverage $0\le \varphi < 2\pi$), or feature partial coverage to suppress contributions from back-to-back jets, as shown in panels (c) and (d) of the same figure.
Panel (a) illustrates a measurement involving two positively charged particles in $A$ and two negatively particles in $B$. A measurement of $B_4^{2+2-}$ shall then be sensitive to the strength (or probability) of processes featuring four correlated particles with two $+$ve and two $-$ve particles separated by a finite rapidity gap.
By contrast, the analysis illustrated in panel (b) would focus on correlated quartets featuring two nearby pairs of $+$ve and $-$ve particles. These could  be produced by string-like fragmentation processes yielding four or more correlated particles, but it could also result from string fragmentation producing two neutral objects, each decaying into pairs of  $+$ve and $-$ve particles.

\section{Multi-particle Balance Functions}
\label{sec:Balance Functions}

Yet another application of integral and differential correlation functions of the form  Eqs.~(\ref{eq:DptDptInt}--\ref{eq:DptDptDptDptInt}), involves the study of (net)-charge (and other quantum numbers) fluctuations. We first show how Eq.~(\ref{eq:DptDptInt}) can be used to measure net-charge fluctuations and how it connects to measurements of balance functions~\cite{Bass:2000az,Pruneau:2022brh,Pruneau:2023zhl}. We also remind the reader how  second moments of the charge are connected to the $\nu_{\rm dyn}$ observable~\cite{Pruneau:2002yf} and differential charge balance functions~\cite{Pruneau:2022brh,Pruneau:2023zhl}. This then provides a convenient mechanism  to introduce higher order balance functions. 

In order to express net-charge fluctuations, we re-write Eq.~(\ref{eq:DptDptInt}) by replacing the transverse momentum $p_{\rm T}$ by the charge $q_i$ of particles.
\begin{equation}
\label{eq:DqDqInt}
\llangle \Delta q_1\Delta q_2 \rrangle \equiv \frac{1}{\langle N(N-1)\rangle}
\int_{\Omega} \Delta q_1\Delta q_2 \rho_2(\vec p_1,\vec p_2) {\rm d}\vec p_1{\rm d}\vec p_2
\end{equation}
where deviates are defined as $\Delta q_i \equiv q_i - \llangle q\rrangle$. The variables $q_i$ are considered  implicit functions of the momentum of the particles and thus cannot be factorized out of the integral. To build this point, let us consider the  expression of the average charge in the acceptance $\Omega$ of interest
\begin{equation}
\label{eq:AvgQ}
\llangle  q \rrangle \equiv \frac{1}{\langle N\rangle}
\int_{\Omega}  \tilde\rho_1(\vec p) {\rm d}\vec p,
\end{equation}
where $\tilde\rho_1(\vec p)$ represents a {\bf  charge density}, i.e., not the   number density $\rho_1(\vec p_1)$. Equation~(\ref{eq:AvgQ}) may be
computed based on number densities if $\tilde\rho_1(\vec p)$ is replaced (symbolically) by  $q_{\alpha} \rho_1^{\alpha}(\vec p)$ where $q_{\alpha}$ is the charge of the particle species of interest, $\alpha$. A similar development can be done for strangeness or baryon quantum numbers by considering strangeness and baryon densities instead of charge densities.   Here for the sake of simplicity, let us restrict the discussion to three types of charged particles:  positively charged, neutral, and negatively charged hadrons. The  single particle density is then $\rho_1(\vec p)=\rho_1^{(+)}(\vec p)+\rho_1^{(0)}(\vec p)+\rho_1^{(-)}(\vec p)$. Substituting this expression in Eq.~(\ref{eq:AvgQ}) and inserting the values $q_i=+1, 0, -1$ for each of the three types, one gets 
\begin{align}
\label{eq:AvgQBis}
\llangle  q \rrangle &\equiv \frac{1}{\langle N\rangle}
\left[ 
(+1) \times \int_{\Omega}\rho_1^{(+)}(\vec p) {\rm d}\vec p + (0) \times  \int_{\Omega}  \rho_1^{(0)}(\vec p) {\rm d}\vec p +  (-1) \times  \int_{\Omega} \rho_1^{(-)}(\vec p) {\rm d}\vec p 
\right] \\  \nonumber
 & = \frac{1}{\left\langle N \right\rangle} 
 \left[ 
 \left\langle N^{+} \right\rangle - \left\langle N^{-} \right\rangle 
\right],
\end{align}
where, in the second line, we applied Eq.~(\ref{eq:f1}). In the following, we  assume neutral particles are not measured and consider the total multiplicity $N$ as the sum of the multiplicities of positively and negatively charged particles, i.e. $N= N^{(+)}+N^{(-)}$. 

The calculation of  $\llangle \Delta q_1\Delta q_2 \rrangle$, which corresponds to the covariance of the charges of two measured particles,  proceeds in the same way. First expand $\Delta q_1\Delta q_2$ and compute its ensemble average
\begin{align}
 \llangle \Delta q_1\Delta q_2 \rrangle  =
 \llangle q_1 q_2 \rrangle - \llangle q \rrangle^2,
\end{align}
which may be expressed according to 
\begin{align}
\llangle \Delta q_1\Delta q_2 \rrangle  =& 
\frac{1}{\langle N(N-1)\rangle}
\iint_{\Omega}
\left[ q_1q_2 \rho_2(\vec p_1,\vec p_2) - q_1 \rho_1(\vec p_1) q_2 \rho_1(\vec p_2) \right] {\rm d}\vec p_1{\rm d}\vec p_2 \\ 
=&\frac{1}{\langle N(N-1)\rangle}\iint_{\Omega} q_1q_2 C_2(\vec p_1,\vec p_2){\rm d}\vec p_1{\rm d}\vec p_2,
\end{align}
where in the second line, we used the expression of the second cumulant $C_2$ given by Eq.~(\ref{eq:DiffCumulant2}). 
Expanding $q \rho_1$ as in Eq.~(\ref{eq:AvgQBis}) and $q_1q_2\rho_2$ according to 
\begin{align}
q_1q_2\rho_2= \rho_2^{(++)} - \rho_2^{(+-)} - \rho_2^{(-+)}+ \rho_2^{(--)},
\end{align}
the integration yields
\begin{align}
\label{eq:DqDqIntBis}
\llangle   \Delta q_1\Delta q_2 \rrangle  =&  
  \frac{  
  \left\langle N^{+}\left(N^{+} - 1\right) \right\rangle
  + \left\langle N^{-} \left(N^{-} - 1 \right) \right\rangle
  - 2\left \langle  N^{+} N^{-}\right\rangle
  - \left( \left\langle N^{+} \right\rangle- \left\langle N^{-} \right\rangle \right)^2}{\langle N(N-1)\rangle} 
  .
\end{align}
At LHC, A--A collisions produces approximately equal multiplicities (and densities) of positively and negatively charged particle: $\langle N^{+} \rangle \approx \langle N^{-} \rangle$.  The above expression for $\llangle   \Delta q_1\Delta q_2 \rrangle$ can thus be written 
\begin{align}
\label{eq:DqDqInt3}
\llangle   \Delta q_1\Delta q_2 \rrangle &\approx \frac{\langle N^{+} \rangle\langle N^{-} \rangle}{\langle N(N-1)\rangle} 
  \left[
  \frac{\langle N^{+} \left(N^{+} - 1  \right) \rangle}{\langle N^{+} \rangle\langle N^{+} \rangle}
  +\frac{\langle N^{-} \left(N^{-} - 1  \right) \rangle}{\langle N^{-} \rangle\langle N^{-} \rangle}
  -2 \frac{\langle N^{+} N^{-}\rangle}{\langle N^{+} \rangle\langle N^{-} \rangle}
  \right]
\end{align}
within which one recognizes  the expression of $\nu_{\rm dyn}^{+-}$~\cite{Pruneau:2002yf}
\begin{align}
\label{eq:NuDyn}
\nu_{\rm dyn}^{+-} &\equiv  
 \frac{\langle N^{+}\left(N^{+}-1 \right) \rangle}{\langle N^{+} \rangle\langle N^{+} \rangle}
  +\frac{\langle N^{-}\left(N^{-}-1 \right) \rangle}{\langle N^{-} \rangle\langle N^{-} \rangle}
  -2 \frac{\langle N^{+} N^{-} \rangle}{\langle N^{+} \rangle\langle N^{-} \rangle} \\ \nonumber
  =& \frac{F_2^{++}}{F_1^+F_1^+} + \frac{F_2^{--}}{F_1^-F_1^-} - 2 \frac{F_2^{+-}}{F_1^+F_1^-} \\ \nonumber
  =& R_2^{++} + R_2^{--} - 2 R_2^{+-},
\end{align}
where in the second line we used the definition of factorial cumulants $F_1^{\alpha}$ and $F_2^{\alpha_1\alpha_2}$ and in the third line, we used the normalized cumulant ratios $R_2^{\alpha_1\alpha_2}$, defined in Eq.~(\ref{eq:DiffNormCumulant2}), with $\alpha_1=+$ and $\alpha_2=-$.
One then obtains the useful result
\begin{align}
\label{eq:DqDqInt4}
\llangle   \Delta q_1\Delta q_2 \rrangle &\approx \frac{\langle N \rangle^2}{4\langle N(N-1)\rangle} 
 \nu_{\rm dyn}^{+-}.
\end{align}  
A similar development can be carried out with a differential version of $\llangle \Delta q_1\Delta q_2 \rrangle$ and one finds 
\begin{align}
\label{eq:DqDqDiff1}
\llangle   \Delta q_1\Delta q_2 \rrangle(\vec p_1,\vec p_2) =& -
\frac{\langle N \rangle}{4\langle N(N-1)\rangle} 
  \left[ B_2^{+-}(\vec p_1,\vec p_2) +  B_2^{-+}(\vec p_1,\vec p_2) \right],
\end{align}
where $B_2^{+-}(\vec p_1,\vec p_2)$ and $B_2^{-+}(\vec p_1,\vec p_2)$ are bound unified balance functions~\cite{Pruneau:2022brh} defined 
according to 
\begin{align}
\label{eq:B2+-Basic}
 B_2^{+-}(\vec p_1,\vec p_2) =& \frac{1}{\langle N^{-} \rangle } \left[ C_2^{+-}(\vec p_1,\vec p_2) - C_2^{--}(\vec p_1,\vec p_2)\right], \\ 
\label{eq:B2-+Basic}
 B_2^{-+}(\vec p_1,\vec p_2) =& \frac{1}{\langle N^{+}  \rangle} \left[ C_2^{-+}(\vec p_1,\vec p_2) - C_2^{++}(\vec p_1,\vec p_2)\right].
\end{align}
The functions $B_2^{+-}(\vec p_1,\vec p_2)$ and $B_2^{-+}(\vec p_1,\vec p_2)$ are constructed in such a way that their 
respective integral each converge to unity in the full acceptance limit~\cite{Pruneau:2022brh}\footnote{for charge conserving processes.}. The fluctuations $\llangle   \Delta q_1\Delta q_2 \rrangle(\vec p_1,\vec p_2)$ thus have an upper bound $\langle N \rangle/\langle N(N-1)\rangle$. And since $\langle N(N-1)\rangle\rightarrow \langle N\rangle^2$ in the large $N$ (and Poisson) limit, one concludes that 
$\llangle   \Delta q_1\Delta q_2 \rrangle(\vec p_1,\vec p_2)$ should scale in inverse proportion of the system size and the produced particle multiplicity. This expectation is verified from a number of recent measurements of charge fluctuations~\cite{Adams:2003st,Abelev:2008jg,Abelev:2012pv}.

It is natural to seek to extend Eq.~(\ref{eq:DqDqDiff1}) to higher moments
by considering expressions of the form $\llangle   \Delta q_1\Delta q_2\, \cdots\, \Delta q_n \rrangle$. We begin, in this section, with a discussion of four-particle balance functions based on an expansion of $\llangle   \Delta q_1\Delta q_2  \Delta q_3 \Delta q_4 \rrangle$.
Extensions to higher orders $n=6, \ldots, 10$ are presented in  Appendix~\ref{sec:Multi-particleBalanceFunctions}. 

In order to compute $\llangle   \Delta q_1\Delta q_2\Delta q_3\Delta q_4 \rrangle$, we first expand the deviates,  compute the event ensemble of the resulting expression and find
\begin{align} \label{eq:Dq1Dq2Dq3Dq4Expression} 
\llangle   \Delta q_1\Delta q_2\Delta q_3\Delta q_4 \rrangle=&
\llangle   q_1 q_2 q_3 q_4 \rrangle
- \llangle q \rrangle 
\llangle q_1 q_2 q_3 + q_1 q_2 q_4 + q_1 q_3 q_4 + q_2 q_3 q_4\rrangle  \\ \nonumber 
&+ \llangle q \rrangle^2
\llangle q_1 q_2 + q_1 q_3 + q_1 q_4 + q_2 q_3 +q_2q_4+q_3q_4\rrangle
- 3\llangle q \rrangle^4.  
\end{align}
We observe the above expression nearly matches the fourth cumulant expansion, Eq.~(\ref{eq:FactorialCumulant4X}), but misses a term of the form $3\llangle \Delta q_1\Delta q_2\rrangle\llangle\Delta q_3\Delta q_4 \rrangle$. This additional term  corresponds to the square of two-particle contributions and needs to be subtracted to eliminate such trivial contributions  to the four particle correlator. Subtracting $3\llangle \Delta q_1\Delta q_2\rrangle\llangle\Delta q_3\Delta q_4 \rrangle$, the integral of $C_4$, denoted $I_4^{+-}$, may then be written according to 
\begin{align}
\label{eq:DqDqDqDqToC4}
I_4^{+-} \equiv  \llangle   \Delta q_1\Delta q_2\Delta q_3\Delta q_4 \rrangle
&-3\llangle \Delta q_1\Delta q_2\rrangle\llangle\Delta q_3\Delta q_4 \rrangle\\
=&
\frac{1}{\langle N(N-1)(N-2)(N-3)\rangle}
\iint_{\Omega}q_1q_2q_3q_4 C_4(\vec p_1, \ldots, \vec p_4) {\rm d}\vec p_1\cdots {\rm d}\vec p_4.
\end{align}
We thus proceed to use the  integral $I_4^{+-}$ and the differential cumulant $C_4(\vec p_1, \ldots, \vec p_4)$  to introduce four-particle balance functions. To this end, we write the charged particle  4-tuplet decomposition of $C_4$ as follows
\begin{align}
\label{eq:4Density}
C_4(\vec p_1,\dots,\vec p_4) =& 
C_4^{++++}(\vec p_1,\dots,\vec p_4) + C_4^{----}(\vec p_1,\dots,\vec p_4) - 4C_4^{+++-}(\vec p_1,\dots,\vec p_4) \\ \nonumber 
& - 4C_4^{---+}(\vec p_1,\dots,\vec p_4) + 6C_4^{++--}(\vec p_1,\dots,\vec p_4),
\end{align}
where we assumed the densities are symmetric under permutations of indices, e.g., $\rho_4^{++--}= \rho_4^{-+-+} == \rho_4^{--++}$, to compute each of the coefficients. The integral, Eq.~(\ref{eq:DqDqDqDqToC4}), becomes
\begin{align}
\label{eq:qqqq}
I_4^{+-} = \frac{1}{N(N-1)(N-2)(N-3))}& \left[ 
\int_{\Omega} C_4^{++++}(\vec p_1,\dots,\vec p_4) {\rm d}\vec p_1\cdots {\rm d}\vec p_4  +\int_{\Omega} C_4^{----}(\vec p_1,\dots,\vec p_4) {\rm d}\vec p_1\cdots {\rm d}\vec p_4 \right. \\ \nonumber 
&- 4\int_{\Omega} C_4^{-+++}(\vec p_1,\dots,\vec p_4) {\rm d}\vec p_1\cdots {\rm d}\vec p_4 
 - 4\int_{\Omega} C_4^{+---}(\vec p_1,\dots,\vec p_4){\rm d}\vec p_1\cdots {\rm d}\vec p_4 \\ \nonumber 
&\left. + 6\int_{\Omega} C_4^{++--}(\vec p_1,\dots,\vec p_4){\rm d}\vec p_1\cdots {\rm d}\vec p_4 \right]\\ 
=& \frac{
\left[  F^{++++}  -4 F^{+++-} + 6 F^{++--} -4 F^{---+}  +  F^{----}\right]
}
{ N(N-1)(N-2)(N-3)
}.
\end{align}
where in the last line we used the definition, Eq.~(\ref{eq:FactorialCumulant4X}), of the 4-particle factorial cumulant.
By analogy to Eq.~(\ref{eq:DqDqDiff1}),  one can then introduce 4-particle differential ``balance functions" according to
\begin{align}
\label{eq:DqDqDqDqVsB4}
I_4^{+-}(\vec p_1,\vec p_2,\vec p_3,\vec p_4) \equiv&
\frac{4!}{2\cdot 2!}
\frac{\langle N^{-}(N^{-}-1)\rangle }{\langle N(N-1)(N-2)(N-3)\rangle} 
B_4^{+-}(\vec p_1,\vec p_2,\vec p_3,\vec p_4)  \\ \nonumber 
&+\frac{4!}{2\cdot 2!}
\frac{\langle N^{+}(N^{+}-1)\rangle }{\langle N(N-1)(N-2)(N-3)\rangle} 
B_4^{-+}(\vec p_1,\vec p_2,\vec p_3,\vec p_4)
\end{align}
where
\begin{align}
\label{eq:b4++--}
B_4^{+-}(\vec p_1,\vec p_2,\vec p_3,\vec p_4) =&  
\frac{2}{4!/2!}
\frac{\left[
 3 C_4^{++--}(\vec p_1,\dots,\vec p_4) 
 - 4 C_4^{+---}(\vec p_1,\dots,\vec p_4) 
 +  C_4^{----}(\vec p_1,\dots,\vec p_4)\right]}
 { \langle N^{-}(N^{-}-1)\rangle }
  \\
\label{eq:b4--++}
B_4^{-+}(\vec p_1,\vec p_2,\vec p_3,\vec p_4) =& 
\frac{2}{4!/2!}
\frac{\left[
 3 C_4^{--++}(\vec p_1,\dots,\vec p_4) 
 - 4 C_4^{-+++}(\vec p_1,\dots,\vec p_4) 
 +  C_4^{++++}(\vec p_1,\dots,\vec p_4)\right]}
 { \langle N^{+}(N^{+}-1)\rangle }.
\end{align}
We use the notations $B_4^{+-}$ and $B_4^{-+}$, 
to denote $n$ particle balance functions involving, the balancing of negatively and positively charged particles by positively and negatively charged particles, respectively.  Inclusion of the ratio of factorial coefficients ($4!/2!$) insures the integral of  $B_4^{\pm\mp}$ converge to unity in the limit of full acceptance. Similar coefficients are introduced, in Appendix~\ref{sec:Multi-particleBalanceFunctions} to achieve proper normalization of the higher order balance functions.
To indeed verify the  integrals of $B_4^{+-}$ and  $B_4^{-+}$ integrate to unity over the full acceptance, let $P(n)$ represent the probability of a process involving the production of $n$ pairs of positively and negative charged particles. Mixed factorial moments are thus trivially given by 
\begin{align}
\label{eq:Average1}
f_1^{\pm} =& \langle N^{\pm}\rangle = \sum_{n=0}^{\infty} n P(n) \equiv  \langle n\rangle , \\ 
\label{eq:Average2LS}
f_2^{\pm\pm} =& \langle N^{\pm}(N^{\pm}-1)\rangle = \langle n^2\rangle - \langle n\rangle, \\ 
\label{eq:Average2US}
f_2^{\pm\mp} =& \langle N^{\pm}N^{\mp}\rangle = \langle n^2\rangle, \\
\label{eq:Average4ls}
f_4^{\pm\pm\pm\pm} =&\langle n^4\rangle - 6\langle n^3\rangle + 11 \langle n^2\rangle - 6\langle n\rangle, \\ 
\label{eq:Average+++-}
f_4^{\pm\pm\pm\mp} =& \langle n^4\rangle - 3\langle n^3\rangle + 2 \langle n^2\rangle, \\ 
\label{eq:Average4++--}
f_4^{\pm\pm\mp\mp} =&  \langle n^4\rangle - 2\langle n^3\rangle + \langle n^2\rangle.
\end{align}
Assuming these are in fact fourth order (or higher) correlations, the integral of $B_4^{+-}(\vec p_1,\dots,\vec p_4)$ over the full momentum volume thus yields
\begin{align}
\label{eq:quads} 
\int_{\Omega} B_4^{+-}(\vec p_1,\dots,\vec p_4) {\rm d}\vec p_1\cdots {\rm d}\vec p_4 = & \frac{2}{4!/2!}
\frac{1}{\langle n(n-1)\rangle}
 \left[
  3\left( \langle n^4\rangle - 2\langle n^3\rangle + \langle n^2\rangle \right) 
  -4 \left( \langle n^4\rangle - 3\langle n^3\rangle + 2 \langle n^2\rangle  \right) \right. \\ \nonumber
  & \left.  +  \left( \langle n^4\rangle - 6\langle n^3\rangle + 11 \langle n^2\rangle - 6\langle n\rangle \right) 
 \right]  = 1
\end{align}
which is indeed equal to unity. By symmetry, the integral of $B_4^{-+}$ also converges to unity in a full acceptance measurement. We  thus have two equivalent four particle balance functions to study charge balanced four particle correlations. 

As for two-particle balance functions, the above expressions can also be generalized to cross species balance function and these shall feature simple sum rules similar to those satisfied by $B_2$ functions~\cite{Pruneau:2022brh}.

Given they are based on 4-particle cumulants, the  functions $B_4^{+-}(\vec p_1,\vec p_2,\vec p_3,\vec p_4)$ and $B_4^{-+}(\vec p_1,\vec p_2,\vec p_3,\vec p_4)$ shall  suppress, by construction, contributions from 4-tuplets of uncorrelated particles. Only 4-tuplets featuring genuinely correlated particles would contribute to the strength of the correlators. Tuplets involving only two or three correlated particles would have vanishing contributions. As such the 4-cumulant components of $B_4$  should indeed suppress contributions from resonance decays resulting into two or three correlated particles. Contributions from hadron resonance  decays would then be limited to four prong decays. Considering these typically have small probabilities, one would expect the magnitude of $B_4$ is then determined by other processes such  as ``string fragmentation", jet fragmentation,   and other multi-particle production processes. Contributions from jet fragmentation processes can be singled out by using a cone-shaped acceptance with, e.g., $\varphi\approx 1$ and $y\approx 1$.  Conversely, jets can be suppressed by using at least two relatively narrow kinematic bins separated by a sizable rapidity gap.  Removal of  two- and three-prong decays, as well as contributions from jets, enables a direct study of the underlying processes, such as string fragmentation, that lead to particle production over extended ranges of rapidity. Long range correlations have already been observed in the context of anisotropic flow measurements  in pp, pA, and AA collisions. These measurements show that the long correlations are  largely dominated by the geometry and fluctuations of the geometry of collisions. They however say little about the underlying nature of the correlations or the correlation length. Measurements of  multi-particle charge (or baryon) balance functions  would change the focus from the transverse geometry to the longitudinal structure of these correlations and might then shed light on the nature and origin of these correlations. 

By construction, $n$-cumulants are non vanishing only if particle correlations of $n$-particle are present in the system considered. Consequently, should observations yield vanishing $4$-cumulants, it would imply that correlation between balancing charges are only limited to second order contributions. If the $4$-cumulants are non-vanishing, they would indicate  more intricate production mechanisms. Either way, measurements of $B_4$  would provide new and valuable information on the structure of particle production dynamics.

Higher order balance functions, $B_n^{\pm\mp}$, with $6 \le n \le 10$ can be constructed in a similar way as $B_4^{\pm\mp}$ and are listed  in Appendix~\ref{sec:Multi-particleBalanceFunctions}. As for $B_4^{\pm\mp}$, higher order balance functions would suppress contributions from lower order correlations. A systematic study of $B_n^{\pm\mp}$ for $n=4$, 6, 8, etc, would then provide sensitivity to increasingly more complicated  production processes featuring a growing number of correlated particles. Such measurements should then provide additional and powerful constrains on multi-particle production models.

\section{Relation to Net-Charge Cumulants}
\label{sec:Net-ChargeCumulants}

Cumulants of the net charge $Q$ (as well as net baryon $B$ and net strangeness $S$) of the particles measured in a specific acceptance $\Omega$ nominally provide 
a probe of the susceptibilities of the matter formed in nucleus-nucleus collisions~\cite{Stephanov:2009ra,Athanasiou:2010kw,Athanasiou:2010vi,Ling:2015yau,Brewer:2018abr}. Several measurements of lower order cumulants (as well as mixed cumulants) have been reported in the recent literature. Ratios of lower order cumulants have been studied, in particular, in the context of the beam energy scans recently performed at the Relativistic Heavy Ion Colliders (RHIC) to  identify signatures of a critical point of nuclear matter~\cite{STAR:2019ans,STAR:2021iop}. Given the potential significance of such critical point, considerable  theoretical and experimental efforts have been deployed to obtain relations between the properties of nuclear matter, net-quantum number cumulants, as well as robust techniques to measure these observables~\cite{Ling:2015yau}.
In this context, note that it was recently shown that a simple relation exists between the second net charge cumulant, $\kappa_2^Q$ and net-charge balance functions $B$~\cite{Pruneau:2019BNC} (also see discussions in \cite{Braun-Munzinger:2019yxj}).  This relation is of particular  interest because it expresses the magnitude of the non trivial part (non-Poissonian) of $\kappa_2^Q$ in terms of an integral of the charge balance function $B$ across a specific experimental acceptance (i.e., a specific kinematic range). Given this integral converges to unity, by construction, in the full acceptance limit, it implies the magnitude of $\kappa_2^Q$  is determined by the shape and width of the balance function relative to the width of the acceptance. This is critical for the beam energy scan because, although the  acceptance can be kept fixed, the shape of the $B$ is known to evolve with produced species, system size, nucleus-nucleus collision centrality, and beam energy~\cite{STAR:2003kbb,Aggarwal:2010ya,Abelev:2013csa,Adamczyk:2015yga,Adam:2015gda,Abelev:2012pv,ALICE:2021hjb}.  As such, the magnitude of $\kappa_2^Q$ thus constitutes a poorly defined reference in the search of a critical point of nuclear matter. That said, it is also of interest to consider how higher cumulants might be impacted by charge conservation, the size of the experimental acceptance of a measurement, and the dynamics of collisions. 

We saw, in sec.~\ref{sec:Balance Functions}, that balance functions naturally arise in the calculation of moments  $\llangle \Delta q_1\Delta q_2\rrangle$ and $\llangle \Delta q_1\Delta q_2\Delta q_3\Delta q_4\rrangle$ and yield expressions proportional to factorial cumulants of the particle multiplicities. We thus seek to determine relations between net-charge cumulants, net-charge factorial cumulants, and factorial cumulants of multiplicities of positively and negatively charge particles.  Details of the derivations are presented in Appendix~\ref{sec:MomentsCumulants}. In this section, we summarize   results of interest which indicate that highest order contributions of net-charge cumulants are identical to integrals of multi-particle balance functions of same order. 

It is well known that moments, cumulants, factorial moments, and factorial cumulants  are readily  computed based on their respective generating functions which, herewith, we denote by $G_m(\theta_Q)$, $G_c(\theta_Q)\equiv\ln G_m(\theta_Q)$, $G_f(s_Q)$, and $G_F(s_Q)\equiv\ln G_f(s_Q)$, where the sub-indices $m$, $c$, $f$, and $F$ indicate generating functions of moments, cumulants, factorial moments, and factorial cumulants, respectively. 
As discussed in Appendix~\ref{sec:NetCharge}, moments $m_n^Q$ are obtained by computing $n$-th derivatives of $G_m$ w.r.t. $\theta_Q$, evaluated at $\theta_Q=0$, while cumulants $\kappa_n^Q$ are obtained from $n$-th order derivatives of  $G_c$ w.r.t. $\theta_Q$ also evaluated at $\theta_Q=0$.  
Given  $G_c= \ln G_m$, it is  then straightforward to compute $\kappa_n^Q$ in terms of moments $m_{n'}^Q$, with $n' \le n$ (See Eqs.~(\ref{eq:kappa1VsM} - \ref{eq:kappa2+-VsM}). Similarly, factorial and factorial cumulants can be also obtained by $n$-th order derivatives of $G_f$ and $G_F$ w.r.t. to $s_Q$. It is however more useful to express the generating functions in terms of multiplicities of positively and negatively charged particles $N_+$ and $N_-$ and their associated dummy variables $\theta_+$ and $\theta_-$ for moments and cumulants calculations and dummy variables $s_+$ and $s_-$ for factorial moments and factorial cumulant calculations. It is then possible to obtain
relations between net-charge cumulants and (mixed) cumulants of $N_+$ and $N_-$ as well as with factorial cumulants of these multiplicities. As shown in detail in Appendix~\ref{sec:NetCharge}, one finds even order $n$-cumulants are given by 
\begin{align}
\label{eq:kappa2^QvsF+-Bis}
\kappa_2^Q =& F_1^+ + F_1^- + F_2^{++} -2 F_2^{+-} + F_2^{--}, \\ 
\label{eq:kappa4^QvsF+-Bis}
\kappa_4^Q =& F_1^{+}     
+ F_1^{-}      
+ \, \cdots \,  
+ F_4^{4+}   
-4 F_4^{3+1-}     
+6 F_4^{2+2-}     
-4 F_4^{1+3-}     
+ F_4^{4-}, \\  
\label{eq:kappa6^QvsF+-Bis}
\kappa_6^Q =& F_1^{+} + F_1^{-} + \, \cdots \,      
+ F_6^{6+} 
-6 F_6^{5+1-} 
+15 F_6^{4+2-} 
-20 F_6^{3+3-} 
+15 F_6^{2+4-} 
-6 F_6^{1+5-} 
+ F_6^{6-}
\end{align}
and so on for higher orders.
Intermediate terms of order $1< n' < n$ were omitted for the sake of clarity.
Comparing the above expressions, as well as Eqs.~(\ref{eq:kappa8^QvsF+-}, \ref{eq:kappa10^QvsF+-}), with Eqs.~(\ref{eq:B2+-} - \ref{eq:B10+-}), we observe that the cumulants $\kappa_{n}^Q$ feature a dependence on the mixed factorial moments $F_n^{k(+)n-k(-)}$ that exactly matches the expression of the balance functions of order $n\ge 4$. Indeed, as for $\kappa_2^Q$, we find  
that the non-trivial component of higher order $\kappa_n^Q$ are exactly proportional to integrals of balance functions $B_{n}^{+-}$, $B_{n}^{-+}$ introduced in sec.~\ref{sec:Balance Functions}. Given these balance functions 
are governed by sum rules, i.e., their full acceptance integrals are entirely determined by charge conservation.  We  conclude that as for second order cumulants $\kappa_2^Q$,
the non-trivial components of higher cumulants, $\kappa_2^Q$, $n \ge 4$, are determined by integral of functions whose full acceptance limit is solely driven by charge conservation. As for basic balance functions, Eqs.~(\ref{eq:B2+-Basic}, \ref{eq:B2-+Basic}), one  expects that these higher order balance function integrals feature a strong dependence on the rapidity and transverse momentum coverage of the measurements, as well as the details of the particle production processes at play in the collisions being studied~\cite{Ling:2015yau}. Additionally, as for basic balance functions, it stands to reason that these higher balance function might feature some dependence on collision centrality and beam energy. Such dependencies might thus be better probed with  differential balance functions.  This suggests that rather than measuring cumulants $\kappa_n^Q$, which only feature information on the integrals of balance functions, it would be
better advised to measure differential balance functions.  Techniques to compute multi-particle balance functions without the drawbacks of multiple nested loops on particles of an event were discussed in sec.~\ref{sec:method} whereas kinematic configurations of measurements of potential interest were presented in sec.~\ref{sec:configurations}.

\section{Summary}
\label{sec:summary}

We first advocated, in sec.~\ref{sec:motivation}, for measurements of integral and differential of multi-particle correlation functions as tools to extract characteristics of heavy ion collisions and the matter they produce heretofore somewhat neglected and susceptible of enhancing the understanding of the physics of these complex systems. We next explicitly presented detailed formula of such multiparticle correlations as well as techniques to compute them in finite time (i.e., single loop on all particles of interest) based on the methods of moments.
This set the stage for the development of what we called multi-particle balance functions. These higher order balance functions were introduced based on expectation values of the form $\llangle \Delta q_1 \Delta q_2 \cdots \Delta q_n\rrangle$ but are best computed in terms of combinations of $n$th order cumulants (or integral factorial cumulants). Much like the original balance functions $B_2$ introduced by Pratt et al.~\cite{Bass:2000az}, these new balance functions are defined in such a way that they integrate to unity in full phase space (i.e., all rapidities and $p_{\rm T}\ge 0$). As such they too provide a measure of the fraction  of charge (or other quantum number) balanced when measured in a finite acceptance. This fraction is expected to be rather sensitive to the details of the (charge conserving) particle production and transport. Indeed, given they are constructed based on $n$-particle cumulants, they should probe the particle production rapidity and momentum correlation length scales and the details of the particle production mechanisms.  

We additionally showed 
these higher order balance functions have integrals, formulated in terms of factorial cumulants, that are equal to the higher order contributions of net charge cumulants $\kappa_n^Q$. This is an important result that pertains to  measurements of net charge (as well as net strangeness and net baryon number) fluctuations  based on cumulants $\kappa_n^Q$ and their evolution with beam energy in the context of the beam energy scan (BES) at the Relativistic Heavy Ion Collider. The magnitude of these cumulants {\bf cannot} be corrected for charge conservation given the balance functions integrate to unity in full acceptance.
Indeed the magnitude of the integral of the balance functions $B_n^{\pm\mp}$ measured in a specific acceptance (in rapidity and transverse momentum) is determined by the details of the particle production and transport (e..g., presence of radial flow) and has thus relatively little to do with the intrinsic properties of the matter they originate from (i.e., the susceptibilities of the QGP). 

The formalism developed in this work for the deployment of multiple particle correlations is in many ways similar to the techniques used in the context of measurements of anisotropic flow. It is thus likely that  the correlation functions discussed in this work might be calculable, with minor or no adaptation,  to existing generic frameworks of anisotropic flow measurements. Of particular interest, however, are practical implementations of $\vec p$ dependent acceptance and efficiency corrections at the single particle level. Also of interest are efficiency losses related to correlated detector effects that likely manifest themselves differently in the context of the correlation functions discussed in this work. The authors thus plan to follow up this work with additional studies of these practical effects.

\newenvironment{acknowledgement}{\relax}{\relax}
\begin{acknowledgement}
	\section*{Acknowledgements}

 The authors thank Drs. S. Pratt, S. Voloshin for insightful discussions and  suggestions.  C.P. also thanks Dr. Gil Paz for technical help with Mathematica.  
 This work  was  supported in part by the United States Department of Energy, Office of Nuclear Physics (DOE NP), United States of America, under grant No.  DE-FG02-92ER40713.  S.B. also acknowledges the support of the Swedish Research Council (VR) and the Knut and Alice Wallenberg Foundation. 

\end{acknowledgement}

\appendix 

\section{Moments, Cumulants, Factorial Moments, and Factorial Cumulants}
\label{sec:MomentsCumulants}

The calculation of moments, cumulants, factorial moments, and factorial cumulants
based on their respective generating functions are discussed in sec.~\ref{sec:Single-VariableSystems} for single variable systems, in sec.~\ref{sec:Multi-VariableSystems} for joint-measurements of multi-variable systems, and in sec.~~\ref{sec:NetCharge} for the specific case of a collision system's net-charge.

\subsection{Single Variable Systems}
\label{sec:Single-VariableSystems}

Recall that given a function $P(N)$ stipulating the probability of observing a value $N$, algebraic moments of $N$, denoted $m_n$, are defined as 
\begin{align}
\label{eq:moments}
m_n \equiv \left\langle N^n\right\rangle = \sum_{N=0}^{\infty} N^n P(N), 
\end{align}
Additionally defining the moment generating function $G_m(\theta) = \langle e^{\theta N}\rangle$,
one readily verifies the moments $m_n$ can be computed according to 
\begin{align}
\label{eq:mVsGm}
m_n  = \left. \partial_{\theta}^n G_m(\theta)\right|_{\theta=0},
\end{align}
where $\partial_{\theta}=\partial/\partial \theta$. Cumulants of $N$ of order $n$, denoted $\kappa_n$, are similarly defined and computed with the introduction of a cumulant generating functions $G_{c}(\theta)\equiv \ln G_m(\theta)$ according to
\begin{align}
\label{eq:cVsGc}
\kappa_n  = \left. \partial_{\theta}^n G_c(\theta)\right|_{\theta=0} = \left. \partial_{\theta}^n \ln G_m(\theta)\right|_{\theta=0}.
\end{align}
Application of the r.h.s. of the above expression readily yields the cumulants $\kappa_n$ as linear combinations of the moments $m_n$. In the context of measurements of particle densities of order $n$, discussed in this work, it is also convenient to consider factorial moments and factorial cumulants. Factorial moments, $f_n$, are formally defined with the introduction of generating functions $G_f(\theta) = \langle s^{N}\rangle$, where $s=e^{\theta}$ and computed according to 
\begin{align}
\label{eq:fVsGf}
f_n  = \left. \partial_{s}^n G_f(s)\right|_{s=1},
\end{align}
where $\partial_{s}=\partial/\partial s$. 
Likewise, factorial cumulants, $F_n$, are formally defined as derivatives of a factorial cumulant generating functions $G_F(s)\equiv\ln G_f(s)$
\begin{align}
\label{eq:FVsGF}
F_n  = \left. \partial_{s}^n G_F(s)\right|_{s=1} = \left. \partial_{s}^n \ln G_f(s)\right|_{s=1}.
\end{align}
Application of the r.h.s. of the above expression yields factorial cumulants $F_n$ as combinations of the factorial moments $f_n$. 
Computation with Mathematica~\cite{Mathematica} yields the following expressions for the ten lowest  orders
\begin{align}
\label{eq:F1vsf}
F_1 =& f_1\\ 
\label{eq:F2vsf}
F_2 =& -f_1^2+f_2\\ 
\label{eq:F3vsf}
F_3 =& 2f_1^3-3f_1f_2 +f_3\\ 
\label{eq:F4vsf}
F_4 =& -6f_1^4-12f_1^2f_2 -3f_2^2-4f_1f_3+f_4\\ 
\label{eq:F5vsf}
F_5 =& 24f_1^5-60f_1^3f_2 +30f_1f_2^2 +20f_1^2f_3-10f_2f_3-5f_1f_4+f_5\\ 
\label{eq:F6vsf}
F_6 =&-120f_1^6+360f_1^4f_2-270f_1^2f_2^2+30f_2^3
-120f_1^3f_3+120f_1f_2f_3-10f_3^2+30f_1^2f_4 \\ \nonumber 
&-15f_2f_4-6f_1f_5+f_6 \\
\label{eq:F7vsf}
F_7=&720f_1^7-2520f_1^5f_2+2520f_1^3f_2^2-630f_1f_2^3+840f_1^4f_3\\ \nonumber
&+ 1260f_1^2f_2f_3 + 210f_2^2f_3+140f_1f_3^2 -210f_1^3f_4+210f_1f_2f_4\\ \nonumber
&-35f_3f_4+42f_1^2f_5-21f_2f_5-7f_1f_6+f_7 \\
\label{eq:F8vsf}
F_8=&-5040f_1^8+20160f_1^6f_2-25200f_1^4f_2+10080f_1^2f_2^3-630f_2^4 \\ \nonumber
&-6720f_1^5f_3+13440f_1^3f_2f_3 - 5040f_1f_2^2f_3 -1680f_1^2f_3^2 \\ \nonumber
&+560f_2f_3^2+1680f_1^4f_4-2520f_1^2f_4+420f_2^2f_4+560f_1f_3f_4-35f_4^2\\ \nonumber
&-336f_1^3f_5+336f_1f_2f_5-56f_3f_5+56f_1^2f_6-28f_2f_6-8f_1f_7+f_8, \\
\label{eq:F9vsf}
F_9=& 40320 f_1^9 - 181440 f_1^7 f_2 + 272160 f_1^5 f_2 - 151200 f_1^3 f_2^3  \\ \nonumber
&+ 22680 f_1 f_2^4 + 60480 f_1^6 f_3 - 151200 f_1 f_2 f_3^2 + 90720 f_1^2 f_2^2 f_3  \\ \nonumber
&- 7560 f_2^3 f_3 + 20160 f_1^3 f_3 - 15120 f_1 f_2 f_3^2 + 560 f_3^3  \\ \nonumber
&- 15120 f_1^5 f_4 + 30240 f_1^3 f_2 f_4 - 11340 f_1 f_2^2 f_4 - 7560 f_1^2 f_3 f_4  \\ \nonumber
&+ 2520 f_2 f_3 f_4 + 630 f_1 f_4^2 + 3024 f_1^4 f_5 - 4536 f_1^2 f_2 f_5  \\ \nonumber
&+ 756 f_2^2 f_5 + 1008 f_1 f_3 f_5 - 126 f_4 f_5 - 504 f_1^3 f_6 + 756 f_2^2 f_5  \\ \nonumber
&+ 1008 f_1 f_3 f_5 - 126 f_4 f_5 - 504 f_1^3 f_6 + 504 f_1 f_2 f_6 - 84 f_3f_6  \\ \nonumber
&+ 72 f_1^2 - 36 f_2f_7 - 9f_1f_8 + f_9,  \\
\label{eq:F10vsf}
F_{10}=&-362880 f_1^{10} + 1814400 f_1^8 f_2 - 3175200 f_1^6 f_2^2 + 2268000 f_1^4 f_2^3 \\ \nonumber
&- 567000 f_1^2f_2^4 + 22680 f_2^5 - 604800 f_1^7 f_3 + 1814400 f_1^5 f_2 f_3 \\ \nonumber
&- 1512000 f_1^3 f_2^2 f_3 + 302400 f_1 f_2^3 f_3 - 252000 f_1^4 f_3^2 \\ \nonumber
&+ 302400 f_1^2 f_2 f_3^2 - 37800 f_2^2 f_3^2 - 16800 f_1 f_3^3 + 151200 f_1^6 f_4 \\ \nonumber
&- 378000 f_1^4 f_2 f_4 + 226800 f_1^2 f_2^2 f_4 - 18900 f_2^3 f_4 + 100800 f_1^3 f_3 f_4 \\ \nonumber
&- 75600 f_1 f_2 f_3 f_4 + 4200 f_3^2 f_4 - 9450 f_1^2 f_4^2 + 3150 f_2 f_4^2 \\ \nonumber
&- 30240 f_1^5 f_5 + 60480 f_1^3 f_2 f_5 - 22680 f_1 f_2^2 f_5 - 15120 f_1^2 f_3 f_5 \\ \nonumber
&+ 5040 f_2 f_3 f_5  + 2520 f_1 f_4 f_5 - 126 f_5^2 + 5040 f_1^4 f_6 - 7500 f_1^2 f_2 f_6 \\ \nonumber
&+ 1260  f_2^2 f_6 + 1680 f_1 f_3 f_6 - 210 f_4 f_6  - 720  f_1^3 f_7  \\ \nonumber
&+ 720 f_1 f_2 f_7 - 120 f_3 f_7 + 90 f_1^2 f_8 - 45 f_2 f_8  - 10 f_1 f_9 + f_{10}   
\end{align}
Factorial cumulants, $F_n$, are of particular interest in the context of measurements of particle production because they identically vanish in the absence of correlations of order $n$.  Note that the relations between cumulants and moments are formally identical to the above given the definitions of cumulants and factorial cumulants in terms of log of their respective generating functions.

It is straightforward (and convenient)  to express cumulants as combinations of  factorial cumulants if one notices that $G_c(\theta)= G_F(s)$ given $s=e^{\theta}$. Taking $n$-order derivatives $\partial_{\theta}^n$ of the l.h.s.  yields  cumulants $\kappa_n$, while derivatives on the r.h.s. are computed based on $\partial_{\theta}= (\partial s/\partial \theta)\partial_{s}=s \partial_{s}$ 
and yield expressions in terms of $F_{n'}$ with $n'\le n$.  The ten lowest orders are 
\begin{align}
\label{eq:kappa1}
    \kappa_1 =& \left. \partial_{\theta} G_c\right|_{\theta=0} = \left. s \partial_{s} G_F\right|_{s=1} = F_1, \\ 
\label{eq:kappa2}
    \kappa_2 =& \left. \partial_{\theta}^2 G_c\right|_{\theta=0} = \left. s \partial_{s} (s \partial_{s} G_F)\right|_{s=1} = F_1 + F_2, \\ 
\label{eq:kappa3}
    \kappa_3 =&F_1+3F_2+F_3, \\
\label{eq:kappa4}
    \kappa_4 =&F_1+7F_2+6F_3+F_4, \\
\label{eq:kappa5}
    \kappa_5 =&F_1+15F_2+25F_3+10F_4+F_5, \\
\label{eq:kappa6}
    \kappa_6 =&F_1+31F_2+90F_3+65F_4+15F_5 +F_6,  \\
\label{eq:kappa7}
    \kappa_7 =&F_1+63F_2+301F_3+350F_4+ 140F_5 + 21F_6 + F_7,  \\
\label{eq:kappa8}
    \kappa_8 =&F_1+127F_2+966 F_3+ 1701 F_4+ 1050 F_5 + 266 F_6 + 28 F_7 + F_8, \\  
\label{eq:kappa9}
    \kappa_9 =&F_1+ 255 F_2+  3025 F_3+ 7770 F_4+ 6951 F_5 + 2646 F_6 + 462  F_7 + 36 F_8 + F_{9}, \\  
\label{eq:kappa10}
 \kappa_{10} =&F_1+ 511 F_2+  9330 F_3+ 34105 F_4+ 45525 F_5 + 22827 F_6 + 5880  F_7 + 750 F_8 + 45 F_{9}+ F_{10}. 
\end{align}
First note that  cumulants of a given order $k$ feature a linear combination 
of all factorial cumulants of lower order $k'\le k$. Second, remember that in the context of particle correlation measurements, one can conclude there are correlations of $k$ or more particles only when a factorial cumulant $F_k$ is non vanishing. Consequently, if a factorial $F_k$ is consistent with zero, within statistical uncertainties, there is no point in measuring $\kappa_k$ or higher order cumulants $\kappa_{k'}$, with $k'>k$ since these do not carry additional experimental information about the system under study. Indeed, in such cases, the magnitude of $\kappa_k$ is primarily determined by factorial cumulants of the lowest orders involving few or, possibly, no particle correlations.

\subsection{Multi-Variable Systems}
\label{sec:Multi-VariableSystems}

Given a function $P(\vec N)$ stipulating the probability of jointly observing $m$ variables $\vec N\equiv (N_1, N_2, \ldots, N_m)$ corresponding to categories $\vec \alpha = (\alpha_1, \alpha_2, \ldots, \alpha_m)$, which in the context of this work corresponds to kinematic bins or particle species or both,  mixed algebraic moments of $\vec N$, denoted $m_{\vec n}^{\vec \alpha}$, are defined as 
\begin{align}
\label{eq:mixedMoments}
m_{\vec n}^{\vec \alpha} \equiv \sum_{\vec N} \prod_{i=1}^m N_i^{n_i} P(\vec N), 
\end{align}
and calculable based on a mixed moment generating functions $G_m(\vec \theta) \equiv \langle e^{\sum_{i=1}^m \theta_i N_i}\rangle$ according to 
\begin{align}
\label{eq:mixmVsGm}
m_{n}^{\vec \alpha}  = \left. \left( \prod_{i=1}^n \partial_{\theta_{i}} \right) G_m(\vec \theta)\right|_{\vec\theta=0},
\end{align}
where $\vec \alpha$ represents all the categories for which moments are evaluated. For instance, a double derivative $\partial_{\theta_1}\partial_{\theta_1}$ would yield a second moment of $N_1$, whereas $\partial_{\theta_1}\partial_{\theta_2}$ would yield a mix moment of $N_1$ and $N_2$.
Proceeding as for a single variable, one defines mixed cumulants 
according to
\begin{align}
\label{eq:mixKappaVsGm}
\kappa_{n}^{\vec \alpha}  = \left. \left( \prod_{i=1}^n \partial_{\theta_{i}} \right)\ln  G_m(\vec \theta)\right|_{\vec\theta=0},
\end{align}
where $\vec\theta=0$ specifies  derivatives are evaluated with $\theta_i=0$, for $i=1,\ldots, m$. Similarly, mixed factorial moments, $f_n^{\vec\alpha}$,  are defined based on mixed moment generating functions $G_f(\vec s) \equiv \langle \prod_{i=1}^m s^{N_i}\rangle$ according to 
\begin{align}
\label{eq:mixfVsGf}
f_{n}^{\vec \alpha}  = \left. \left( \prod_{i=1}^n \partial_{s_{i}} \right) G_f(\vec s)\right|_{\vec s=1},
\end{align}
where $\vec s=1$ specifies  derivatives are evaluated with $s_i=1$, for $i=1,\ldots, m$. Factorial cumulants are defined as derivatives of the factorial cumulant generating functions $G_F(\vec s) \equiv \ln G_f(\vec s)$
according to 
\begin{align}
\label{eq:mixFVsGf}
F_{n}^{\vec \alpha}  = \left. \left( \prod_{i=1}^n \partial_{s_{i}} \right) \ln G_f(\vec s)\right|_{\vec s=1},
\end{align}
Mathematica~\cite{Mathematica} enables a speedy and reliable computation of $F_{n}^{\vec \alpha}$. The lowest orders are found to be
 \begin{align}
 \label{eq:FactorialCumulant2X}
  F_2^{\alpha_1\alpha_2} =& f_2^{\alpha_1\alpha_2}  
  - f_1^{\alpha_1} f_1^{\alpha_2} \\  
 \label{eq:FactorialCumulant3X}
  F_3^{\alpha_1\cdots\alpha_3} =& f_3^{\alpha_1\cdots\alpha_3}  
  - \sum_{(3)} f_2^{\alpha_1\alpha_2} f_1^{\alpha_3} + 2 f_1^{\alpha_1}  f_1^{\alpha_2}f_1^{\alpha_3}\\  
 \label{eq:FactorialCumulant4X}
  F_4^{\alpha_1\cdots\alpha_4} =& f_4^{\alpha_1\cdots\alpha_4}  
  - \sum_{\rm (4)} f_3^{\alpha_1\cdots\alpha_3} f_1^{\alpha_4}  
  - \sum_{\rm (3)} f_2^{\alpha_1\alpha_2} f_2^{\alpha_3\alpha_4}  + 2 \sum_{\rm (6)} f_2^{\alpha_1\alpha_2} f_1^{\alpha_3}f_1^{\alpha_4}
 - 6 f_1^{\alpha_1} \times \cdots \times f_1^{\alpha_4}, \\ 
 \label{eq:FactorialCumulant5X}
  F_5^{\alpha_1\cdots\alpha_5} =& f_5^{\alpha_1\cdots\alpha_5}  
  - \sum_{\rm (5)} f_4^{\alpha_1\cdots\alpha_4} f_1^{\alpha_5}  
  - \sum_{\rm (10)} f_3^{\alpha_1\cdots\alpha_3} f_2^{\alpha_4\alpha_5}  - \sum_{\rm (10)} f_3^{\alpha_1\cdots\alpha_3} f_1^{\alpha_4}f_1^{\alpha_5} + 2 \sum_{\rm (15)} f_2^{\alpha_1\alpha_2} f_2^{\alpha_3\alpha_4}f_1^{\alpha_5}  \\ \nonumber
  &-6  \sum_{\rm (10)} f_2^{\alpha_1\alpha_2} f_1^{\alpha_3}f_1^{\alpha_4}f_1^{\alpha_5}
 +24f_1^{\alpha_1} \times \cdots \times f_1^{\alpha_5}, \\
\label{eq:FactorialCumulant6X}
  F_6^{\alpha_1\cdots\alpha_6} =& f_6^{\alpha_1\cdots\alpha_6}  
  - \sum_{\rm (6)} f_5^{\alpha_1\cdots\alpha_5} f_1^{\alpha_6}  
  - \sum_{\rm (15)} f_4^{\alpha_1\cdots\alpha_4} f_2^{\alpha_5\alpha_6} - \sum_{\rm (15)} f_4^{\alpha_1\cdots\alpha_4} f_1^{\alpha_5} f_1^{\alpha_6}  \\ \nonumber
  &- \sum_{\rm (10)} f_3^{\alpha_1\alpha_2\alpha_3} f_3^{\alpha_4\alpha_5\alpha_6} 
  +2 \sum_{\rm (60)} f_3^{\alpha_1\alpha_2\alpha_3} f_2^{\alpha_4\alpha_5} f_1^{\alpha_6} 
  - 6\sum_{\rm (20)} f_3^{\alpha_1\alpha_2\alpha_3} f_1^{\alpha_4} f_1^{\alpha_5}f_1^{\alpha_6}  \\ \nonumber
  &+ 2 \sum_{\rm (15)} f_2^{\alpha_1\alpha_2} f_2^{\alpha_3\alpha_4}f_2^{\alpha_5\alpha_6} 
  + -6 \sum_{\rm (45)} f_2^{\alpha_1\alpha_2} f_2^{\alpha_3\alpha_4}f_1^{\alpha_5}f_1^{\alpha_6} 
  + 24 \sum_{\rm (15)} f_2^{\alpha_1\alpha_2} f_1^{\alpha_3}f_1^{\alpha_4}f_1^{\alpha_5}f_1^{\alpha_6} \\ \nonumber 
  & - 120 f_1^{\alpha_1} \times \cdots \times f_1^{\alpha_6},
\end{align}
where the notation $\sum_{\rm (k)}$ stands for a sum over $k$ (ordered) permutations of the labels $\alpha_1$, $\alpha_2$, and $\alpha_3$, $\ldots$.

As in the case of a single variable, one can express mixed cumulants $\kappa_{n}^{\vec \alpha}$ in terms of factorial cumulants 
$F_{n}^{\vec \alpha}$ based on the equality $G_c(\vec \theta) = G_F(\vec s)$
by taking derivatives on l.h.s. relative to $\theta_i$ whereas derivative are taken relative to $\partial_{\theta_i}= (\partial s_i/\partial \theta_i)\partial_{s_i}=s_i \partial_{s_{i}}$ on the r.h.s..

\subsection{Net Charge \texorpdfstring{$Q$}\xspace}
\label{sec:NetCharge}

Let $Q=N_+ - N_-$ and $N=N_+ + N_-$ define the net-charge and total charged particle multiplicity, respectively, detected in a given event, with $N_+$ and $N_-$ respectively representing the number of positively and negatively charged particles in that event. The number of positively and negatively charged particles are expected to fluctuate on an event-by-event basis both owing the stochastic nature of the particle production and variations in the processes yielding particles. 
Moments and cumulants of $Q$  are of interest because they nominally relate to charge susceptibility of the matter formed in heavy-ion collisions~\cite{Athanasiou:2010kw,Stephanov:2009ra,Brewer:2018abr,Athanasiou:2010vi}. Moments $m_n^Q$ of the net-charge are defined, as in Eq.~(\ref{eq:moments}), according to 
\begin{align}
m_n^Q \equiv \left\langle Q^n\right\rangle = \sum_{Q} Q^n P(Q,N), 
\end{align}
where $P(Q,N)$ represents the probability of observing a net-charge $Q$ and total multiplicity $N$ in a particular event.
Moments $m_n^Q $ can evidently be computed based on a generating function of the form $G_m(N,Q) = \langle e^{N \theta_N + Q \theta_Q} \rangle$ but it is of greater interest to obtain the moments, the cumulants, and so on, in terms of moments of the multiplicities $N_+$ and $N_-$. Clearly, a simple change of variable enables the definition of  $P(N_+,N_-)$ as the joint probability of observing events with 
$N_+$ and $N_-$ positively and negatively charged particles. The moment generating functions 
of (mixed) moments of the multiplicities can then be written
\begin{align}
G_m(\theta_+,\theta_-) = \langle  e^{N_+ \theta_+ + N_- \theta_-} \rangle
\end{align}
and successive derivatives of $G_m$ w.r.t. $\theta_+$ and $\theta_-$, evaluated at $\theta_+=\theta_-=0$ yield moments and mixed moments of $N_+$ and $N_-$. Introducing the notations $\partial_+ =\partial/\partial\theta_+$ and $\partial_- =\partial/\partial\theta_-$, one computes lowest order mixed moments according to
\begin{align}
m_1^{\pm} =& \left.\partial_{\pm} G_m(\theta_+,\theta_-)\right|_{\theta_+=\theta_-=0}  = \langle N_{\pm}\rangle, \\ 
m_2^{\pm\pm} =& \left.\partial_{\pm}\partial_{\pm} G_m(\theta_+,\theta_-)\right|_{\theta_+=\theta_-=0} =  \langle N_{\pm}^2\rangle, \\ 
m_2^{+-} =& \left.\partial_{-}\partial_{+} G_m(\theta_+,\theta_-)\right|_{\theta_+=\theta_-=0} =  \langle N_{+}N_{-}\rangle,
\end{align}
and so on. Cumulants  and mixed cumulants  of the multiplicities $N_+$ and $N_-$ are computed based on the cumulant generating function $G_c(\theta_+,\theta_-)\equiv \ln G_m(\theta_+,\theta_-)$ by taking successive derivatives w.r.t. $\theta_+$ and $\theta_-$. One for instance gets
\begin{align}
\label{eq:kappa1VsM}
\kappa_1^{\pm} =& 
\left. \partial_{\pm} G_c(\theta_+,\theta_-)\right|_{\theta_+=\theta_-=0} = \left. G_m^{-1}\partial_{\pm} G_m \right|_{\theta_+=\theta_-=0}  = \langle N_{\pm}\rangle, \\ 
\label{eq:kappa2VsM}
\kappa_2^{\pm\pm} =& 
\left. \partial_{\pm} \left(G_m^{-1}\partial_{\pm} G_m \right)\right|_{\theta_+=\theta_-=0} =\langle N_{\pm}^2\rangle - \langle N_{\pm}\rangle^2, \\ 
\label{eq:kappa2+-VsM}
\kappa_2^{+-} =& \left. \partial_{-} \left(G_m^{-1}\partial_{+} G_m \right)\right|_{\theta_+=\theta_-=0} =  \langle N_{+}N_{-}\rangle - \langle N_{+}\rangle\langle N_{-}\rangle, 
\end{align}
and similarly for higher orders. One recognizes $\kappa_2^{\pm\pm}$ and $\kappa_2^{+-}$ as the variance and covariance of  $N_+$ and $N_-$ while higher order $\kappa_3$ and $\kappa_4$ (not shown) are related to skewness and kurtosis of these multiplicities. 

In the context of measurements of net charge fluctuations, it is of interest to relate cumulants of the net charge $Q$ to factorial cumulants of the $N_+$ and $N_-$. First note that  the factorial moments are calculable based on the factorial moment generating functions defined 
as  $G_f(s_+,s_-) \equiv \left\langle s_+^{N_+} s_-^{N_-}\right\rangle $, where $s_+=e^{\theta_+}$ and $s_-=e^{\theta_-}$.  Introducing the notations 
$\partial_{s_+} \equiv \partial/\partial s_+$
and $\partial_{s_-}\equiv \partial/\partial s_-$,  mixed factorial moments of $N_+$ and $N_-$ are obtained by repeated evaluations of derivatives  $\partial_{s_+}$
and $\partial_{s_-}$ evaluated at $s_+=s_-=1$.
\begin{align}
f_1^{\pm} =& 
\left. \partial_{s_\pm} G_f(s_+,s_-)\right|_{s_+=s_-=1} =
\left. \langle  N_{\pm} s_{\pm}^{N_{\pm}-1} s_{\mp}^{N_{\mp}}\rangle\right|_{s_+=s_-=1} = \langle N_{\pm}\rangle, \\ 
f_2^{\pm\pm} =& \left. \partial_{s_\pm} \langle  N_{\pm} s_{\pm}^{N_{\pm}-1} s_{\mp}^{N_{\mp}}\rangle\right|_{s_+=s_-=1} = \langle N_{\pm}(N_{\pm}-1)\rangle, \\ 
f_2^{+-} =& \left. \partial_{s_-} \langle  N_+ s_+^{N_+-1} s_-^{N_-}\rangle\right|_{s_+=s_-=1} = \langle N_{+}N_{-}\rangle, \\ 
f_3^{\pm\pm\pm} =&\langle N_{\pm}(N_{\pm}-1)(N_{\pm}-2)\rangle, \\ 
f_3^{\pm\pm\mp} =&\langle N_{\pm}(N_{\pm}-1)N_{\mp}\rangle, 
\end{align}
and so on. In order to compute the relations between $\kappa_n^Q$ and factorial cumulants of the multiplicities $N_+$ and $N_-$, we introduce  $\partial_{\theta_Q}$, with $\theta_{Q} = \theta_{+} - \theta_{-}$, as a linear combination  of $\partial_{\theta_+}$ and $\partial_{\theta_-}$ according
\begin{align}
    \partial_{\theta_Q} \equiv \frac{\partial}{\partial \theta_Q} 
    = \frac{\partial \theta_+}{\partial \theta_Q}\frac{\partial}{\partial \theta_+} + \frac{\partial \theta_-}{\partial \theta_Q}\frac{\partial}{\partial \theta_-} 
    = \partial_{\theta_+} - \partial_{\theta_-}.
\end{align}
Cumulants of $Q$ of order $n$ are  obtained by computing $n$ derivatives of $G_c(\theta_Q,\theta_N)$  w.r.t. $\theta_{Q}$ according to 
\begin{align}
    \kappa_n^Q \equiv & \partial_{\theta_Q}^n G_c(\theta_Q,\theta_N) = \left( \partial_{\theta_+} - \partial_{\theta_-} \right)^n G_c(\theta_+,\theta_-) \\ \nonumber 
    =& \sum_{k=0}^n  \left(-\right)^{n-k}  \binom{n}{k} 
    \partial_{\theta_+}^k \partial_{\theta_-}^{n-k} G_c(\theta_+,\theta_-) 
      = \sum_{k=0}^n  \left(-\right)^{n-k}  \binom{n}{k}\kappa_n^{(k)(n-k)}, 
\end{align}
where $\kappa_n^{(k)(n-k)}$ represent mixed cumulants of order $k$ and $n-k$ in $N_+$ and $N_-$. One gets 
\begin{align}
\label{eq:kappa1vsKappa+-}
    \kappa_1^Q =& \kappa_1^+ - \kappa_1^-, \\
\label{eq:kappa2vsKappa+-}
    \kappa_2^Q =& \kappa_2^{2+} - 2 \kappa_2^{1+1-} + \kappa_2^{2-}, \\ 
\label{eq:kappa3vsKappa+-}
    \kappa_3^Q =& \kappa_3^{3+} - 3 \kappa_3^{2+1-} + 3 \kappa_3^{1+2-} - \kappa_3^{3-},
\end{align}
and so on. 
Experimentally, it is of greater interest to obtain the cumulants  $\kappa_n^Q$ in terms of factorial cumulants because these are easier to correct for (single) particle losses and vanish in the absence of correlations at order $n$. Evidently, it is only meaningful to report cumulants $\kappa_n^Q$ if the corresponding factorial cumulant $F_n$ are non-vanishing since only these provide new
information not already included in cumulants of lower orders.
Replacing derivatives $\partial_{\theta_i}$ by $s_i \partial_{s_i}$, and noting that $\partial s_i/\partial \theta_j = \delta_{ij} s_i \partial_{s_i}$, one gets
\begin{align}
\label{eq:kappaNn^Qgeneric}
    \kappa_n^Q =& \sum_{k=0}^n  \left(-1\right)^{n-k}  \binom{n}{k}  
    \left( s_+ \partial_{s_+} \right)^k \left( s_-\partial_{s_-}\right)^{n-k} G_F(s_+,s_-).
\end{align}
Lowest orders of interest for this work are found to be 
\begin{align}
\label{eq:kappa1^QvsF+-}
\kappa_1^Q =& F_1^+ - F_1^-, \\ 
\label{eq:kappa2^QvsF+-}
\kappa_2^Q =& F_1^+ + F_1^- + F_2^{2+} -2 F_2^{+-} + F_2^{2-}, \\ 
\label{eq:kappa3^QvsF+-}
\kappa_3^Q =& F_1^+ - F_1^- - 3F_2^{2+}  - 3F_2^{2-} + F_3^{3-} +3F_3^{1+2-} -+3F_3^{2+1-}  - F_3^{3+}
, \\     
\label{eq:kappa4^QvsF+-}
\kappa_4^Q =& F_1^{+}     
+ F_1^{-}     
+7 F_2^{2+}    
-2 F_2^{+-}     
+ 7 F_2^{2-}  +6 F_3^{3+}     
-6 F_3^{1+2-}     
-6 F_3^{2+1-}     
+ 6 F_3^{3-} \\   \nonumber
&+ F_4^{4+}   
-4 F_4^{3+1-}     
+6 F_4^{2+2-}     
-4 F_4^{1+3-}     
+ F_4^{4-}, \\  
\label{eq:kappa5^QvsF+-}
\kappa_5^Q =& F_1^{+} - F_1^{-}     
+15 F_2^{2+}    
-15 F_2^{2-} 
+25 F_3^{3+}     
-15 F_3^{2+1-}     
+15 F_3^{1+2-}     
-25 F_3^{3-} \\     \nonumber
&
+10 F_4^{4+}   
-20 F_4^{3+1-}
+20 F_4^{1+3-}     
-10 F_4^{4+0-} \\      \nonumber
&+   F_5^{5+}
-5 F_5^{4+1-}    
+10 F_5^{3+2-}    
-10 F_5^{2+3-}    
+5 F_5^{1+4-}    
-  F_5^{5-} \\ 
\label{eq:kappa6^QvsF+-}
\kappa_6^Q =& F_1^{+} + F_1^{-}     
+ 31 F_2^{2+} 
- 2 F_2^{1+1-} 
+ 31 F_2^{2-} 
+ 90 F_3^{3+} 
- 30 F_3^{2+1-} 
- 30 F_3^{1+2-} 
+ 90 F_3^{3-} \\    \nonumber 
&+ 65 F_4^{4+} 
-80 F_4^{3+1-} 
+ 30 F_4^{2+2-} 
-80 F_4^{1+3-} 
+ 65 F_4^{4-} \\ \nonumber 
&+ 15 F_5^{5+} 
-45 F_5^{4+1-} 
+30 F_5^{3+2-} 
+30 F_5^{2+3-} 
-45 F_5^{1+4-} 
+ 15 F_5^{5-}  \\ \nonumber 
&+ F_6^{6+} 
-6 F_6^{5+1-} 
+15 F_6^{4+2-} 
-20 F_6^{3+3-} 
+15 F_6^{2+4-} 
-6 F_6^{1+5-} 
+ F_6^{6-} 
\\
\label{eq:kappa8^QvsF+-}
\kappa_8^Q =& F_1^{+} + F_1^{-}
+127  F_2^{2+}  
- 2   F_2^{1+1-}  
+ 127 F_2^{2-}  \\ \nonumber
&+966  F_3^{3+}  
-126  F_3^{1+2-}  
-126  F_3^{2+1-}  
+ 966 F_3^{3-}   \\ \nonumber
&+1701  F_4^{4+}  
-924   F_4^{3+1-}  
+126   F_4^{2+2-}  
-924   F_4^{1+3-}  
+ 1701 F_4^{4-}  \\ \nonumber
&+1050  F_5^{5+}  
-1470  F_5^{4+1-}  
+420   F_5^{3+2-}  
+420   F_5^{2+3-}  
-1470  F_5^{1+4-}  
+1050  F_5^{5-}  \\ \nonumber
&+266 F_6^{6+}  
+756 F_6^{1+5-}  
+630 F_6^{2+4-}  
-280 F_6^{3+3-}  
+ 266 F_6^{6-}  
+630 F_6^{4+2-}  
-756 F_6^{5+1-}  \\ \nonumber
&+ 28 F_7^{7+}  
- 140 F_7^{6+1-}  
+ 252 F_7^{5+2-}  
- 140 F_7^{4+3-}  
- 140 F_7^{3+4-}  \\ \nonumber
&+ 252 F_7^{2+5-}  
- 140 F_7^{1+6-}  
+ 28 F_7^{7-}  \\ \nonumber
&+   F_8^{8+}
-8  F_8^{7+1-}  
+28 F_8^{6+2-}  
-56 F_8^{5+3-}  
+70 F_8^{4+4-}   \\ \nonumber
&-56 F_8^{3+5-}  
+28 F_8^{2+6-}  
-8 F_8^{1+7-}  
+  F_8^{8-}  \\
\label{eq:kappa10^QvsF+-}
\kappa_{10}^Q =& F_1^{+} + F_1^{-} +\, \cdots\, + \\  \nonumber
&+    F_{10}^{10+}
-10   F_{10}^{9+1-}  
+45   F_{10}^{8+2-}  
-120  F_{10}^{7+3-}  
+210  F_{10}^{6+4-}  
-252  F_{10}^{5+5-}  \\  \nonumber
&+210 F_{10}^{4+6-}  
-120  F_{10}^{3+7-}  
+45   F_{10}^{2+8-} 
+10   F_{10}^{1+9-} 
+     F_{10}^{10-},
\end{align}
where, for $\kappa_{10}^Q$, we omitted terms of lesser interest.
We first note that cumulants $\kappa_n^Q$ of order $n$ feature a dependence on factorial cumulants of all orders $n'\le n$. Also recall, once again, that factorial cumulants of order $n'$ are non vanishing if and only if $n'$ or more particles are correlated in the events of interest. This implies that high order cumulants $\kappa_n^Q$ can be non-vanishing  based on single particle or correlated pairs even in the absence of $n$ particle correlations. 
Higher order cumulants, $n\ge 3$, are thus non-trivial, i.e., carry new information relative to lower orders,  only if factorial cumulants of same order are non vanishing. 

Additionally note that $\kappa_2^Q$ depends on $F_2^{++} -2 F_2^{+-} + F_2^{--}$ which 
amounts to the integral of the two-particle balance function $B_2^{+-}$ across the acceptance of the measurement. Similarly, one observes that  $\kappa_4^Q$ depends on 
$F_4^{4+} -4 F_4^{3+1-} +6 F_4^{2+2-} -4 F_4^{1+3-} + F_4^{4-}$ which corresponds to the average of the four-particle balance functions, $\frac{1}{2}(B_4^{+-}+B_4^{+-})$ we introduced in sec.~\ref{sec:Multi-particleBalanceFunctions}. Additional inspection of the expressions for higher (even) order net-charge cumulants $\kappa_n^Q$ reveal these also contain sums of mixed charged particle cumulants corresponding to the higher order balance functions defined in Appendix~\ref{sec:Multi-particleBalanceFunctions}. 
As shown in that Appendix, given  the integrals of multi-particle balance functions  are constrained by sum rules determined by charge conservation, we conclude that the magnitude of the cumulants $\kappa_n^Q$, for even values of $n$, are also entirely determined by effects associated to charge conservation and the widths of the measurement acceptance. 
A comprehensive study of the cumulants $\kappa^Q$ with beam energy and system size thus requires 
a detailed understanding of the evolution of the factorial cumulants $F_n$ with beam energy and system size. Given it is likely that multi-particle balance functions of produced particles have intricate dependencies on beam energy, and in particular the growing impact of nuclear stopping at lower energy, we advocate that differential measurements of balance functions provide better insight in the impact of effects associated to the collision dynamics that may otherwise impede the studies of the properties being sought for.

\section{Differential Correlations}
\label{sec:DifferentialCorrelations}
Differential correlation functions of $n$ particles, herein simply termed  $n$-cumulants, may be obtained at any order $n$ by listing 
all distinct ways to ``cluster" $n$ particles into smaller subsets (i.e., clusters) in order to obtain $n$-particle densities in terms of correlated clusters of particles  (and thus cumulants) of lower order $n'\le n$. Cumulants of order $n\le 4$ can then be written
\begin{align}
\label{eq:DiffCumulant1X}
C_1^{\alpha}(\vec p) \equiv &   \rho_1^{\alpha}(\vec p) \\
\label{eq:DiffCumulant2X}
C_2^{\alpha_1\alpha_2}(\vec p_1,\vec p_2) \equiv &   \rho_2^{\alpha_1\alpha_2}(\vec p_1,\vec p_2) 
- C_1^{\alpha_1}(\vec p_1)C_1^{\alpha_2}(\vec p_2) \\
\label{eq:DiffCumulant3X}
C_3^{\alpha_1\cdots\alpha_3}(\vec p_1,\ldots,\vec p_3) \equiv &   \rho_3^{\alpha_1\cdots\alpha_3}(\vec p_1,\ldots,\vec p_3)
- \sum_{\rm (3)}C_2^{\alpha_1\alpha_2}(\vec p_1,\vec p_2)C_1^{\alpha_3}(\vec p_3) 
- C_1^{\alpha_1}(\vec p_1)C_1^{\alpha_2}(\vec p_2)C_1^{\alpha_3}(\vec p_3) \\
\label{eq:DiffCumulant4X}
\nonumber 
C_4^{\alpha_1\cdots\alpha_4}(\vec p_1,\ldots,\vec p_4) \equiv &  
\rho_4^{\alpha_1\cdots\alpha_4}(\vec p_1,\ldots,\vec p_4) 
- \sum_{\rm (4)}C_3^{\alpha_1\cdots\alpha_3}(\vec p_1,\vec p_2,\vec p_3)C_1^{\alpha_4}(\vec p_4) 
- \sum_{\rm (3)}C_2^{\alpha_1\alpha_2}(\vec p_1,\vec p_2)C_2^{\alpha_3\alpha_4}(\vec p_3,\vec p_4)   \\ 
&- \sum_{\rm (6)} C_2^{\alpha_1\alpha_2}(\vec p_1,\vec p_2)C_1^{\alpha_3}(\vec p_3)C_1^{\alpha_4}(\vec p_4)
-C_1^{\alpha_1}(\vec p_1)C_1^{\alpha_2}(\vec p_2)C_1^{\alpha_3}(\vec p_3)C_1^{\alpha_4}(\vec p_4),
%
\end{align}
where the symbols $\rho_n$ and $C_n$ respectively indicate $n$-particle densities and cumulants at momenta $\vec p_1$, $\ldots$, $\vec p_n$ while the labels $\alpha_1, \ldots, \alpha_n$ are species or kinematic bins identifiers. 
Clearly, a formula for $C_2^{\alpha_1\alpha_2}(\vec p_1,\vec p_2)$ in terms of densities is readily obtained by substituting expressions for  $C_1$ by single densities $\rho_1^{\alpha}(\vec p)$. Similarly, and recursively, cumulants of order $n\ge 2$ can  be computed based on lower order cumulants $n'<n$. Alternatively, one can also formulate a generating function $G_{\rho}$ according to~\cite{Kitazawa:2017ljq}
\begin{align}
    G_{\rho}[\theta_1(\vec p_{\rm T}),\ldots, \theta_m(\vec p_{\rm T})] \equiv & \int \exp\left\{\int \prod_{i=1}^m \theta_i(\vec p) n_i(\vec p)\right\} P[n(\vec p_{\rm T})]{\it D}n
\end{align}
where $\theta_1, \ldots, \theta_m$ identify $m$ species or types of particles, $n_i(\vec p)$ is the density of particles of type $i$, and $P[n(\vec p_{\rm T})]$ is the probability of having particles of type $i$ with densities $n(\vec p_{\rm T})$ on an event-by-event basis, and 
$\int  P[n(\vec p_{\rm T})]{\it D}n=1$. The moments of the densities $n_i$ are then obtained in the usual way by computing derivatives w.r.t. $\theta_i(\vec p)$ evaluated at $\theta_i(\vec p)=0$.
\begin{align}
    \langle n_{\alpha_1}(\vec p_1)\, \cdots\,  n_{\alpha_n}(\vec p_n)\rangle 
    \equiv& \int n_{\alpha_1}(\vec p_1)\, \cdots\, n_{\alpha_n}(\vec p_n)P[n(\vec p_{\rm T})]{\it D}n \\ \nonumber
   =& \left. \partial_{\theta_1(\vec p_1)}\, \cdots\,  \partial_{\theta_n(\vec p_n)} G_{\rho}(\theta_1(\vec p_1), \ldots, \theta_n(\vec p_1) \right|_{\theta_i=0}.
\end{align}
Of interest also are factorial moments and factorial cumulants corresponding to these moments. These are obtained by introducing  continuous variables $s_i(\vec p_i)= \exp[\theta_i(\vec p_i)]$ and expressing  $G_{\rho}$ as a function of these variables
\begin{align}
    G_{\rho}[s_1(\vec p_{\rm T}),\ldots, s_m(\vec p_{\rm T})] \equiv & \int \exp\left\{\int \prod_{i=1}^m n_i(\vec p) \ln s_i(\vec p) \right\} P[n(\vec p_{\rm T})]{\it D}n.
\end{align}
Factorial cumulants, corresponding to connected  parts (i.e., correlated) of the densities, are then obtained by taking functional derivatives of $\ln G_{\rho}[s_1(\vec p_{\rm T}),\ldots, s_m(\vec p_{\rm T})]$ w.r.t. $s_i(\vec p_i)$. Structurally, expressions of $n$-cumulants in terms of densities $\rho_n$ have the same dependence as cumulants $F_n$ on factorial moments $f_n$, one can use the relations (\ref{eq:F1vsf}-\ref{eq:F10vsf}) and substitute densities $\rho_k$ to moments $f_k$, with $k=0,\ldots, n$ to obtain the connected correlation functions of interest. 

Additionally note, as already stated in sec.~\ref{sec:cumulants}, that integrals of densities $\rho_n^{\alpha_1\cdots\alpha_n}(\vec p_1,\ldots,\vec p_n)$ yield factorial moments
\begin{align}
\label{eq:factorialMoments}
    f_n^{\alpha_1\cdots\alpha_n} \equiv& \int\cdots\int_{\Omega} \rho_n^{\alpha_1\cdots\alpha_n}(\vec p_1,\ldots,\vec p_n) {\rm d}\vec p_1 \cdots {\rm d}\vec p_n 
    =   \langle N(N-1)\, \cdots\, (N-n+1) \rangle,
\end{align}
whereas integrals of correlation functions  $C_n^{\alpha_1\cdots\alpha_n}(\vec p_1,\ldots,\vec p_n)$ yield factorial cumulants 
\begin{align}
\label{eq:factorialCumulants}
    F_n^{\alpha_1\cdots\alpha_n}  \equiv& \int\cdots\int_{\Omega} C_n^{\alpha_1\cdots\alpha_n}(\vec p_1,\ldots,\vec p_n) {\rm d}\vec p_1 \cdots {\rm d}\vec p_n.
\end{align}
There is indeed a one-to-one relation between densities $\rho_k(\vec p_1,\ldots, \vec p_k)$ and factorial moments $f_k$ as well as between functional cumulants (or cumulant  densities) $C_k(\vec p_1,\ldots, \vec p_k)$ and factorial cumulants $F_k$.

\section{Event ensemble average of deviates}
\label{sec:DeviatesAverage}

Expressing event ensemble average of deviates $\llangle \Delta q_1 \, \cdots \,\Delta q_n\rrangle$ in terms of sums
averages $\llangle q_1 \, \cdots \,  q_m\rrangle$, $m\le n$, is a trivial but somewhat tedious process. We  created simple scripts to  compute these for arbitrary orders $n$ and show below expressions up to order 6. 
\begin{align}
\label{eq:Dq1Dq2}
\llangle \Delta q_1 \Delta q_2\rrangle =& \llangle q_1q_2\rrangle  - \llangle q\rrangle ^{2}, \\ 
\label{eq:Dq1Dq2Dq3}
\llangle \Delta q_1 \Delta q_2\Delta q_3\rrangle =& 
\llangle q_1q_2q_3\rrangle  
- 3\llangle q\rrangle \llangle q_1q_2\rrangle  
+2 \llangle q\rrangle ^{3} \\ 
\label{eq:Dq1Dq2Dq3Dq4}
\llangle \Delta q_1  \, \cdots \,\Delta q_4\rrangle =&
\llangle q_1q_2q_3q_4\rrangle  
-4 \llangle q\rrangle \llangle q_1q_2q_3\rrangle  
+6 \llangle q\rrangle ^{2}\llangle q_1q_2\rrangle  
-3 \llangle q\rrangle ^{4}  \\
\label{eq:Dq1Dq2Dq3Dq4Dq5}
\llangle \Delta q_1 \, \cdots \,\Delta q_5\rrangle =& 
\llangle q_1 \, \cdots \,q_5\rrangle  
- 5\llangle q\rrangle \llangle q_1q_2q_3q_4\rrangle  
+10 \llangle q\rrangle ^{2}\llangle q_1q_2q_3\rrangle  
-10 \llangle q\rrangle ^{3}\llangle q_1q_2\rrangle +4 \llangle q\rrangle ^{5} \\ 
\label{eq:Dq1Dq2Dq3Dq4Dq5Dq6}
\llangle \Delta q_1  \, \cdots \,\Delta q_6\rrangle =& 
+\llangle q_1 \, \cdots \,q_6\rrangle  
- 6\llangle q\rrangle \llangle q_1 \, \cdots \,q_5\rrangle  
+15 \llangle q\rrangle ^{2}\llangle q_1q_2q_3q_4\rrangle  \\ \nonumber
&-20 \llangle q\rrangle ^{3}\llangle q_1q_2q_3\rrangle  
+15 \llangle q\rrangle ^{4}\llangle q_1q_2\rrangle  
-5 \llangle q\rrangle ^{6} 
\end{align}
Inspection  of the above expressions reveal a simple pattern based on binomial  coefficients as follows
\begin{align}
\label{eq:Dq1-Dqm}
\llangle \Delta q_1  \, \cdots \,\Delta q_{n}\rrangle =& \sum_{k=0}^n  (-1)^{n-k} \binom{n}{k} \llangle q\rrangle^{n-k} \llangle  q_{1}\,\cdots\,q_{k} \rrangle, 
\end{align}
where we define $\llangle  q_{1}\,\cdots\,q_{k} \rrangle \equiv \llangle q\rrangle$ for $k=1$  and $\llangle  q_{1}\,\cdots\,q_{k} \rrangle \equiv 1$ for $k=0$.

Moments of cross deviates of variable $q$ and $p$ are computed in a similar fashion. One gets at lowest orders
\begin{align} 
\label{eq:Delta_q(1) Delta_p(1)}
\llangle  \Delta q_1 \Delta  p_1\rrangle =&  \llangle q_1 p_1\rrangle - \llangle q\rrangle\llangle p\rrangle, \\ 
\label{eq:Delta_q(2) Delta_p(1)}
\llangle  \Delta q_1 \Delta q_2 \Delta p_1\rrangle =& \llangle q_1q_2p_1\rrangle  
- \llangle p\rrangle \llangle q_1q_2\rrangle  
- 2\llangle q\rrangle \llangle q_1p_1\rrangle  
+2 \llangle p\rrangle  \llangle q\rrangle ^{2} \\ 
\label{eq:Delta_q(3) Delta_p(1)}
\llangle  \Delta q_1 \Delta q_2 \Delta q_3 \Delta p_1\rrangle =&
\llangle q_1q_2q_3p_1\rrangle  
- \llangle p\rrangle \llangle q_1q_2q_3\rrangle  
-3 \llangle q\rrangle \llangle q_1q_2p_1\rrangle  
+3 \llangle p\rrangle  \llangle q\rrangle \llangle q_1q_2\rrangle \\ \nonumber
&+3 \llangle q\rrangle ^{2}\llangle q_1p_1\rrangle  
-3 \llangle p\rrangle  \llangle q\rrangle ^{3} \\ 
\label{eq:Delta_q(4) Delta_p(1)}
\llangle  \Delta q_1 \, \cdots \, \Delta q_4 \Delta p_1\rrangle =&
\llangle q_1\, \cdots \,q_4p_1\rrangle 
- \llangle p\rrangle \llangle q_1\, \cdots \,q_4\rrangle  
-4 \llangle q\rrangle \llangle q_1q_2q_3p_1\rrangle 
+4 \llangle p\rrangle  \llangle q\rrangle \llangle q_1q_2q_3\rrangle   \\ \nonumber
&+6 \llangle q\rrangle ^{2}\llangle q_1q_2p_1\rrangle  
-6 \llangle p\rrangle  \llangle q\rrangle ^{2}\llangle q_1q_2\rrangle  
-4 \llangle q\rrangle ^{3}\llangle q_1p_1\rrangle  
+4 \llangle p\rrangle  \llangle q\rrangle ^{4} \\
\label{eq:Delta_q(2) Delta_p(2)}
\llangle  \Delta q_1 \Delta q_2 \Delta p_1 \Delta p_1\rrangle =&\llangle q_1q_2p_1p_2\rrangle  
-2 \llangle p\rrangle \llangle q_1q_2p_1\rrangle  
+ \llangle p\rrangle ^{2}\llangle q_1q_2\rrangle  
-2 \llangle q\rrangle \llangle q_1p_1p_2\rrangle  \\ \nonumber
&+4 \llangle p\rrangle  \llangle q\rrangle \llangle q_1p_1\rrangle  
-3 \llangle p\rrangle ^{2} \llangle q\rrangle ^{2} 
+ \llangle q\rrangle ^{2}\llangle p_1p_2\rrangle\\
\label{eq:Delta_q(3) Delta_p(2)}
\llangle  \Delta q_1 \, \cdots \, \Delta q_3 \Delta p_1 \Delta p_2\rrangle =&\llangle q_1q_2q_3p_1p_2\rrangle  
-2 \llangle p\rrangle \llangle q_1q_2q_3p_1\rrangle  
+ \llangle p\rrangle ^{2}\llangle q_1q_2q_3\rrangle  
-3 \llangle q\rrangle \llangle q_1q_2p_1p_2\rrangle  \\ \nonumber
&+6 \llangle p\rrangle  \llangle q\rrangle \llangle q_1q_2p_1\rrangle  
-3 \llangle p\rrangle ^{2} \llangle q\rrangle \llangle q_1q_2\rrangle  
+3 \llangle q\rrangle ^{2}\llangle q_1p_1p_2\rrangle  \\ \nonumber
&-6 \llangle p\rrangle  \llangle q\rrangle ^{2}\llangle q_1p_1\rrangle  
+4 \llangle p\rrangle ^{2} \llangle q\rrangle ^{3} 
- \llangle q\rrangle ^{3}\llangle p_1p_2\rrangle.
\end{align} 
Once again, close inspection of the above expressions reveal straightforward patterns and one obtains a generic formula in terms of two binomial coefficients as follows
\begin{align} 
\label{eq:Delta_q(n) Delta_p(m)}
\llangle \Delta q_1  \, \cdots \,\Delta q_{n}\Delta p_1  \, \cdots \,\Delta p_{m} \rrangle =& \sum_{k=0}^n \sum_{l=0}^m (-1)^{n-k}(-1)^{m-l} \binom{n}{k}\binom{m}{l} \llangle q\rrangle^{n-k} 
\llangle p\rrangle^{m-l} 
\llangle  q_{1}\,\cdots\,q_{k} \rrangle, 
\llangle  p_{1}\,\cdots\,p_{l} \rrangle, 
\end{align} 
Similarly, products of $\Delta q$s, $\Delta p$s, and $\Delta r$s yield at lowest orders
\begin{align} 
\label{eq:Delta_q(1) Delta_p(1) Delta_r(1)}
\llangle \Delta  q_i \Delta p_j \Delta r_k\rrangle =& 
\llangle  q_i p_j r_k\rrangle
-\llangle p\rrangle\llangle  q_i r_j\rrangle
-\llangle q\rrangle\llangle  p_i r_j\rrangle  
-\llangle r\rrangle\llangle  q_i p_j\rrangle 
+2\llangle q\rrangle\llangle  p\rrangle\llangle  r\rrangle \\ 
\label{eq:Delta_q(2) Delta_p(1) Delta_r(1)}
\llangle  \Delta q_i \Delta q_j \Delta p_k \Delta r_l\rrangle =& 
\llangle  q_i q_j p_k r_l\rrangle 
-2 \llangle q\rrangle\llangle  q_i p_j r_k\rrangle
+ \llangle q\rrangle^2\llangle  p_i r_j\rrangle
- \llangle p\rrangle\llangle  q_i q_j r_k\rrangle
+2 \llangle q\rrangle\llangle p\rrangle \llangle  q_i r_j\rrangle \\ \nonumber
&-3 \llangle q\rrangle^2\llangle p\rrangle \llangle  r\rrangle
- \llangle r\rrangle\llangle q_iq_jp_k\rrangle
+2 \llangle q\rrangle\llangle r\rrangle \llangle  q_i p_j\rrangle
+\llangle p\rrangle\llangle r\rrangle \llangle  q_i q_j\rrangle \\ 
\label{eq:Delta_q(2) Delta_p(1) Delta_r(1) Delta_s(1)}
\llangle  \Delta q_i \Delta p_j \Delta r_k \Delta s_l\rrangle =& 
\llangle  q_i p_j r_k s_l\rrangle 
- \llangle p\rrangle\llangle  q_i r_j s_k\rrangle
- \llangle q\rrangle\llangle  p_i r_j s_k\rrangle
- \llangle p\rrangle\llangle  q_i q_j r_k\rrangle \\ \nonumber
&+ \llangle q\rrangle\llangle p\rrangle \llangle  r_i s_j\rrangle 
- \llangle r\rrangle\llangle q_ip_js_l\rrangle
+ \llangle p\rrangle\llangle r\rrangle\llangle q_is_j\rrangle
+\llangle s\rrangle\llangle  q_i p_jr_k\rrangle 
+\llangle p\rrangle\llangle s\rrangle\llangle  q_i r_j\rrangle \\ \nonumber
&+\llangle q\rrangle\llangle s\rrangle\llangle  p_i r_j\rrangle 
+\llangle r\rrangle\llangle s\rrangle\llangle  q_i p_j\rrangle 
+3 \llangle q\rrangle\llangle p\rrangle\llangle r\rrangle \llangle  s\rrangle.
\end{align}
In this case, inspection of the above expressions reveals a formula involving three binomial coefficients 
\begin{align} 
\label{eq:Delta_q(n) Delta_p(m) Delta_r(o)}
\llangle 
\Delta q_1  \, \cdots \,\Delta q_{n}
\Delta p_1  \, \cdots \,\Delta p_{m} 
\Delta r_1  \, \cdots \,\Delta r_{o} 
\rrangle =& \sum_{i=0}^n \sum_{j=0}^m\sum_{k=0}^l (-1)^{n-i}(-1)^{m-k}(-1)^{o-k} 
\binom{n}{i}\binom{m}{j}\binom{o}{k}  \\ \nonumber
& \times \llangle q\rrangle^{n-i} 
\llangle p\rrangle^{m-j} 
\llangle p\rrangle^{o-k} 
 \times \llangle  q_{1}\,\cdots\,q_{i} \rrangle
\llangle  p_{1}\,\cdots\,p_{j} \rrangle
\llangle  r_{1}\,\cdots\,r_{l} \rrangle. 
\end{align} 
Similar formula are readily obtained for  four or more variables.

\section{Computation of Correlators \texorpdfstring{$\llangle q_1^{n_1} q_2^{n_2}\,\cdots\,q_m^{n_m}\rrangle$}\xspace}
\label{sec:computation}

In section sec.~\ref{sec:method}, we derived expressions for   $\llangle q_1^{n_1} q_2^{n_2}\rrangle$
and  $\llangle q_1^{n_1} q_2^{n_2}q_3^{n_3}\rrangle$ in terms of functions of the event-wise sums $Q_n$ defined by Eq.~(\ref{eq:avg Qn}). The same approach can be used to define higher moments of the form $\llangle q_1^{n_1} \, \cdots \, q_m^{n_m}\rrangle$, for arbitrary values of $m$. First note that the products of more than three $Q$s can be computed by straightforward expansion of the sums corresponding to each variable $Q_n$. At order $m$, one obtains  expressions of the form
\begin{align}
\label{eq:Q_n1 Q_n2 Q_nm} \\
Q_{n_1} Q_{n_2} \,\cdots\,Q_{n_m} =&  \sum_{i=1}^N q_i^{n_1+n_2+\cdots+n_m}  
+ \sum_{\rm perms}\sum_{i_1\ne i_2=1}^N q_{i_1}^{n_1+n_2+\cdots+n_{m-1}}q_{i_2}^{n_m} \\ \nonumber
&+ \sum_{\rm perms} \sum_{i_1\ne i_2=1}^N q_{i_1}^{n_1+n_2+\cdots+n_{m-2}}q_{i_2}^{n_{m-1}+n_m} + \sum_{\rm perms} \sum_{i_1\ne i_2\ne i_3=1}^N q_{i_1}^{n_1+n_2+\cdots+n_{m-2}}q_{i_2}^{n_{m-1}}q_{i_3}^{n_m} \\ \nonumber
&+ \sum_{\rm perms}  \sum_{i_1\ne i_2=1}^N q_{i_1}^{n_1+n_2+\cdots+n_{m-3}}q_{i_2}^{n_{m-2}+n_{m-1}+n_m} 
+ \sum_{\rm perms} \sum_{i_1\ne i_2\ne i_3=1}^N q_{i_1}^{n_1+n_2+\cdots+n_{m-3}}q_{i_2}^{n_{m-2}+n_{m-1}}q_{i_3}^{n_{m}} \\ \nonumber
&+ \sum_{\rm perms} \sum_{(i_1,\cdots,i_m)=1}^N q_{i_1}^{n_1+n_2+\cdots+n_{m-3}}q_{i_2}^{n_{m-2}} q_{i_3}^{n_{m-1}} q_{i_4}^{n_{m}} +\,\,  \cdots \,\, \\ \nonumber
 &+ \sum_{\rm perms} \sum_{(i_1,\cdots,i_m)=1}^N q_{i_1}^{n_1+n_2} q_{i_2}^{n_{3}} \,\cdots\,  q_{i_{m-2}}^{n_{m-1}} q_{i_{m-1}}^{n_{m}} + \sum_{(i_1,\cdots,i_m)=1}^N q_{i_1}^{n_1}q_{i_2}^{n_2}  \,\cdots\,q_{i_m}^{n_m},
\end{align}
where the notation $\sum_{\rm perms}$ indicates a sum over all ordered  permutations of the exponents $n_i$, $i=1,\ldots, m$, whereas $\sum_{(i_1,\cdots,i_m)=1}^N$ represents a sum over all distinct $m$-tuples of values of the indices $i_1, i_2, \ldots, i_m$, i.e., $i_1\ne i_2\ne \cdots \ne i_m$.  When an ensemble average is computed, each term of the above expression yields terms of the form $\sum_{\rm perms} \langle N(N-1)\cdots(N-m+1)\rangle \llangle q_1^{n_1}\, \cdots\,q_m^{n_m}\rrangle$. Clearly, terms of the form $\llangle q_1^{n_1}\, \cdots\,q_m^{n_m}\rrangle$ can be computed iteratively based on sums of  $\llangle q_1^{n_1}\, \cdots\,q_p^{n_p}\rrangle$, with $p\le m$. It is thus nominally possible to obtain expressions for $\llangle q_1^{n_1}\, \cdots\,q_m^{n_m}\rrangle$ at any order $m$ based on sums of lower order terms. In practice, one finds that the number of terms to be considered grows very rapidly as $m$ increases. We thus opted to write scripts (in C{\small{++}}) based on the TString root class~\cite{Brun:1997pa}. The computation proceeds in four basic steps. In the first step, at given order $m$, one  finds all the permutations of exponents $n_1$, $n_2$, etc, that yield terms that are products of two factors  $q_1^aq_2^b$, three factors $q_1^aq_2^bq_3^c$, and so on. Once these permutations are listed, one proceeds to generate these terms by recursively calling functions that generate them from lower order products. This second step is then followed by 
an aggregation and simplification step in which identical terms are regrouped and rearranged to produce a latex output. Low orders were checked against the results of manual calculations and low order $m\le 4$ expressions published elsewhere for $n_1=n_2=\cdots=n_m=1$~\cite{Giacalone:2020lbm,ALICE:2023tej}.  As an example, we show below the expression for arbitrary integer values $n_1, \cdots, n_m$ obtained for $m=4$:
\begin{align}
  \langle N(N-1)\,\cdots\,(N-3)\rangle \llangle  q_1^{n_{1}} q_2^{n_{2}}q_3^{n_{3}}q_4^{n_{4}} \rrangle =& 
-6 \llangle Q_{n1+n2+n3+n4}\rrangle 
+2 \llangle Q_{n1+n2+n3}Q_{n4}\rrangle 
+2 \llangle Q_{n1+n2+n4}Q_{n3}\rrangle \\ \nonumber 
&+2 \llangle Q_{n1+n3+n4}Q_{n2}\rrangle 
+2 \llangle Q_{n2+n3+n4}Q_{n1}\rrangle 
+\llangle Q_{n1+n2}Q_{n3+n4}\rrangle \\ \nonumber 
&+\llangle Q_{n1+n3}Q_{n2+n4}\rrangle 
+\llangle Q_{n1+n4}Q_{n2+n3}\rrangle 
-\llangle Q_{n1+n2}Q_{n3}Q_{n4}\rrangle \\ \nonumber 
&-\llangle Q_{n1+n3}Q_{n2}Q_{n4}\rrangle
-\llangle Q_{n1+n4}Q_{n2}Q_{n3}\rrangle  
-\llangle Q_{n2+n3}Q_{n1}Q_{n4}\rrangle \\ \nonumber
&-\llangle Q_{n2+n4}Q_{n1}Q_{n3}\rrangle
-\llangle Q_{n3+n4}Q_{n1}Q_{n2}\rrangle 
+\llangle Q_{n1}Q_{n2}Q_{n3}Q_{n4}\rrangle \\ \nonumber
=& -6 \llangle Q_{n_1+n_2+n_3+n_4}\rrangle 
  +2 \sum_{(4)} \llangle Q_{n_1+n_2+n_3}Q_{n_4}\rrangle  \\ \nonumber 
&  +\sum_{(3)}\llangle Q_{n_1+n_2}Q_{n_3+n_4}\rrangle  
 -\sum_{(6)}\llangle Q_{n_1+n_2}Q_{n_3}Q_{n_4}\rrangle \\ \nonumber 
 & +\llangle Q_{n_1}Q_{n_2}Q_{n_3}Q_{n_4}\rrangle,  
\end{align}
where  the notation $\sum_{(n)}$ indicate sums over all ordered permutations of the indices $n_1$, $n_2$, $n_3$, and $n_4$~\footnote{The code used to generate these and other expressions reported in this paper  is available on Github (https://github.com/cpruneau/MomentsCalculator.git).}.

The computation of event ensemble averages of  deviates, Eqs.~(\ref{eq:Dq1Dq2} - \ref{eq:Dq1-Dqm}), require expressions for products of the form $\llangle q_1\, \cdots \,q_m\rrangle$. These are obtained by setting exponents $n_1=n_2=\,\cdots\,n_m=1$ in the generic expressions $\llangle q_1^{n_1}\, \cdots \,q_n^{n_m}\rrangle$. Although somewhat simpler, these remain fastidious to calculate by hand. We have extended our  scripts to automatically set exponents $n_i$, $i=1,\ldots, m$  to unity programmatically. Computation of the first eight orders yields
\begin{align}
\label{eq:q1q2VsQs}
 \langle N(N-1)\rangle \llangle  q_1 \cdots q_2 \rrangle =& 
 -\llangle Q_{2}\rrangle   +\llangle Q_{1}^{2}\rrangle  \\
\label{eq:q1q2q3VsQs}
 \langle N(N-1)(N-2)\rangle \llangle  q_1 \cdots q_3 \rrangle =& 
 2\llangle Q_{3}\rrangle -3 \llangle Q_{2}Q_{1}\rrangle  +\llangle Q_{1}^{3}\rrangle \\
\label{eq:q1q2q3q4VsQs}
 \langle N(N-1) \, \cdots \,(N-3)\rangle \llangle  q_1 \cdots q_4 \rrangle =& 
 -6 \llangle Q_{4}\rrangle +8 \llangle Q_{3}Q_{1}\rrangle +3 \llangle Q_{2}^{2}\rrangle -6 \llangle Q_{2}Q_{1}^{2}\rrangle +\llangle Q_{1}^{4}\rrangle \\
 \label{eq:q1q2q3q4q5VsQs}
 \langle N(N-1)\,\cdots\,(N-4)\rangle \llangle  q_1 \, \cdots \,q_5 \rrangle =&
 24 \llangle Q_{5}\rrangle 
-30 \llangle Q_{4}Q_{1}\rrangle 
-20 \llangle Q_{2}Q_{3}\rrangle 
+20 \llangle Q_{3}Q_{1}^{2}\rrangle 
+15 \llangle Q_{2}^{2}Q_{1}\rrangle \\ \nonumber 
&-10 \llangle Q_{2}Q_{1}^{3}\rrangle 
+\llangle Q_{1}^{5}\rrangle \\
\label{eq:q1q2q3q4q5q6VsQs}
 \langle N(N-1)\,\cdots\,(N-5)\rangle \llangle  q_1 \, \cdots \,q_6 \rrangle =&
-120 \llangle Q_{6}\rrangle 
+144 \llangle Q_{5}Q_{1}\rrangle 
+90 \llangle Q_{4}Q_{2}\rrangle   
-90 \llangle Q_{4}Q_{1}^{2}\rrangle
+40 \llangle Q_{3}^{2}\rrangle    \\ \nonumber
&-120 \llangle Q_{3}Q_{2}Q_{1}\rrangle 
+40 \llangle Q_{3}Q_{1}^{3}\rrangle 
-15 \llangle Q_{2}^{3}\rrangle 
+45 \llangle Q_{2}^{2}Q_{1}^{2}\rrangle 
-15 \llangle Q_{2}Q_{1}^{4}\rrangle  \\ \nonumber 
&+\llangle Q_{1}^{6}\rrangle \\
\label{eq:q1q2q3q4q5q6q7VsQs}
 \langle N(N-1) \, \cdots \,(N-6)\rangle \llangle  q_1 \cdots q_7 \rrangle =& 
+530 \llangle Q_{7}\rrangle 
-850 \llangle Q_{6}Q_{1}\rrangle 
-294 \llangle Q_{5}Q_{2}\rrangle 
+504 \llangle Q_{5}Q_{1}^{2}\rrangle   \\ \nonumber
&-335 \llangle Q_{4}Q_{3}\rrangle 
+630 \llangle Q_{4}Q_{2}Q_{1}\rrangle 
-210 \llangle Q_{4}Q_{1}^{3}\rrangle    \\ \nonumber
&+290 \llangle Q_{3}^{2}Q_{1}\rrangle 
+105 \llangle Q_{3}Q_{2}^{2}\rrangle 
-420 \llangle Q_{3}Q_{2}Q_{1}^{2}\rrangle   \\ \nonumber
&+ 70 \llangle Q_{3}Q_{1}^{4}\rrangle 
-105 \llangle Q_{2}^{3}Q_{1}\rrangle 
+105 \llangle Q_{2}^{2}Q_{1}^{3}\rrangle   \\ \nonumber
&-21 \llangle Q_{2}Q_{1}^{5}\rrangle 
+\llangle Q_{1}^{7}\rrangle
\\
\label{eq:q1q2q3q4q5q6q7q8VsQs}
 \langle N(N-1) \, \cdots \,(N-7)\rangle \llangle  q_1 \cdots q_8 \rrangle =& 
-4760 \llangle Q_{8}\rrangle 
+7720 \llangle Q_{7}Q_{1}\rrangle 
+2450 \llangle Q_{6}Q_{2}\rrangle   \\ \nonumber
&-3430 \llangle Q_{6}Q_{1}^{2}\rrangle 
+3528 \llangle Q_{5}Q_{3}\rrangle 
-5712 \llangle Q_{5}Q_{2}Q_{1}\rrangle   \\ \nonumber
&+1344 \llangle Q_{5}Q_{1}^{3}\rrangle 
+770  \llangle Q_{4}^{2}\rrangle 
-4480 \llangle Q_{4}Q_{3}Q_{1}\rrangle   \\ \nonumber
&-630  \llangle Q_{4}Q_{2}^{2}\rrangle 
+2520 \llangle Q_{4}Q_{2}Q_{1}^{2}\rrangle 
-420  \llangle Q_{4}Q_{1}^{4}\rrangle   \\ \nonumber
&-1470 \llangle Q_{3}^{2}Q_{2}\rrangle 
+1190 \llangle Q_{3}^{2}Q_{1}^{2}\rrangle 
+2520 \llangle Q_{3}Q_{2}^{2}Q_{1}\rrangle   \\ \nonumber
&-1120 \llangle Q_{3}Q_{2}Q_{1}^{3}\rrangle 
+112  \llangle Q_{3}Q_{1}^{5}\rrangle 
+105 \llangle Q_{2}^{4}\rrangle   \\ \nonumber
&-420 \llangle Q_{2}^{3}Q_{1}^{2}\rrangle 
+210 \llangle Q_{2}^{2}Q_{1}^{4}\rrangle 
-28  \llangle Q_{2}Q_{1}^{6}\rrangle  
+\llangle Q_{1}^{8}\rrangle.
\end{align}
Formula for the ensemble average of deviates of the form 
$\llangle \Delta q_1\, \cdots\, \Delta q_m \rrangle$, shown in Eqs.~(\ref{eq:Dq1Dq2}--\ref{eq:Dq1-Dqm}), 
are obtained by substitution of the expressions for $\llangle q_1\, \cdots\, q_m \rrangle$ listed above. The three lowest orders are  
\begin{align}
 \label{eq:Dq1Dq2VsQ}
 \llangle \Delta q_1 \Delta q_2 \rrangle  =& 
 \frac{\llangle Q_1^2 \rrangle -\llangle Q_2 \rrangle }{\langle N(N-1)\rangle}
 -  \frac{\llangle Q_1\rrangle^2}{\langle N\rangle^2}, \\
 \label{eq:Dq1Dq2Dq3VsQ}
 \llangle \Delta q_1 \Delta q_2\Delta q_3 \rrangle  =& 
 \frac{\llangle Q_1^3 \rrangle  -3\llangle Q_2Q_1 \rrangle + 2\llangle Q_3 \rrangle}{\langle N(N-1)(N-2)\rangle} 
-3 \frac{\llangle Q_1\rrangle}{\langle N\rangle}
 \frac{\left(\llangle Q_1^2 \rrangle-\llangle Q_2 \rrangle \right)}{\langle N(N-1)\rangle}
+ 2 \frac{\llangle Q_1\rrangle^3}{\langle N\rangle^3}, \\
\label{eq:Dq1Dq2Dq3Dq4VsQ}
 \llangle \Delta q_1 \Delta q_2\Delta q_3\Delta q_4 \rrangle  =& 
\frac{ 
\llangle Q_{1}^{4} \rrangle
+3 \llangle Q_{2}^{2}\rrangle 
-6 \llangle Q_{2}Q_{1}^{2}\rrangle 
+8 \llangle Q_{3}Q_{1}\rrangle 
-6 \llangle Q_{4}\rrangle 
}{
\langle N(N-1)(N-2)(N-3)\rangle
} \\ \nonumber 
&-4  \frac{\llangle Q_1\rrangle}{\langle N\rangle}
 \frac{\left( \llangle Q_1^3 \rrangle  -3\llangle Q_2Q_1 \rrangle + 2\llangle Q_3 \rrangle\right)}{\langle N(N-1)(N-2)\rangle} 
+6 \frac{\llangle Q_1\rrangle^2}{\langle N\rangle^2}
\frac{\left(\llangle Q_1^2 \rrangle -\llangle Q_2 \rrangle\right) }{\langle N(N-1)\rangle}
-3\frac{\llangle Q_1\rrangle^4}{\langle N\rangle^4}
\end{align}
The above expressions of event ensemble averages of products of deviates $\Delta q_i$ involve inclusive averaging, i.e., computation of the average of products and powers of $Q$s separately. These are then divided by averages of the multiplicity $\langle N\rangle$ and average numbers of $n$-tuplets $\langle N(N-1)\ldots (N-n+1)\rangle$. One can readily switch to event-wise averaging, corresponding to calculations of the products and powers on an event-by-event basis by ``moving" the double brackets to include the divisions by the number of $n$-tuples. For instance, the lowest two orders may be written~\cite{Giacalone:2020lbm}
\begin{align}
 \label{eq:Dq1Dq2VsQeventwise}   
 \llangle \Delta q_1 \Delta q_2 \rrangle  =& 
 \left\llangle \frac{Q_1^2  - Q_2}{ N(N-1)}\right\rrangle
 - \left\llangle \frac{ Q_1}{N}\right\rrangle^2, \\
 \label{eq:Dq1Dq2Dq3VsQeventwise}
 \llangle \Delta q_1 \Delta q_2\Delta q_3 \rrangle  =& 
 \left\llangle
 \frac{Q_1^3 -3 Q_2Q_1+ 2 Q_3}{N(N-1)(N-2)}
 \right\rrangle 
-3 \left\llangle\frac{Q_1}{N}\right\rrangle
 \left\llangle\frac{\left(Q_1^2-Q_2\right)}{N(N-1)}\right\rrangle
+2 \left\llangle \frac{Q_1}{N}\right\rrangle^3, 
\end{align}
where the notation $\llangle R\rrangle$ denote ensemble averaging of ratios, $R$, of
 functions of $Q$s, calculating event-by-event, and the number of $n$-tuples formed by the $N$ particles of a given event.

The computation of ensemble averages of  mixed moment deviates based on event-wise sums of variables $q_i$, $p_i$, $r_i$, $s_i$, $t_i$, etc,  proceeds in a similar fashion.  One first defines event-wise sums $Q_n$, $P_n$, $R_n$, $S_n$, $T_n$ according to
\begin{align}
    Q_n =& \sum_{i=1}^N q_i^n, \hspace{0.3in} 
    P_n = \sum_{i=1}^N p_i^n, \hspace{0.3in}  
    R_n = \sum_{i=1}^N r_i^n, \hspace{0.3in}
    S_n = \sum_{i=1}^N s_i^n, \hspace{0.3in}  
    T_n = \sum_{i=1}^N t_i^n,\hspace{0.3in}\rm etc.
\end{align}
One next lists all required mixed products of these sums and finally proceed to evaluate their event ensemble averages. 

We limit the discussion to three variables, $q_i$, $p_j$, $r_k$, corresponding to three distinct kinematic bins or species, but the technique is readily applicable to an arbitrary number of such variables.
We thus seek to express cross moments of interest, $\llangle q_1\cdots q_{m_1}p_1\cdots p_{m_2}r_1\cdots r_{m_3}\rrangle$, in terms of ensemble averages of products of $Q_n$, $P_n$, and $R_n$, as appropriate. 
Given the moments defined in Eqs.~(\ref{eq:Delta_q(1) Delta_p(1)} - \ref{eq:Delta_q(n) Delta_p(m) Delta_r(o)}),
one expects to need  ensemble averages of the form  $\llangle Q_n\rrangle$, $\llangle Q_nQ_m\rrangle$, $\llangle Q_nQ_mQ_o\rrangle$ already computed in sec.~\ref{sec:method} (and in this appendix) as well as  cross moments of the form 
$\llangle Q_nP_m\rrangle$, $\llangle Q_nQ_mP_o\rrangle$, $\llangle Q_nQ_mQ_oP_p\rrangle$, $\llangle Q_nQ_mP_oP_p\rrangle$, etc, that we now proceed to compute.  All other cross moments can be obtained  by appropriate permutations of variable names and indices.  Let $N_q$, $N_p$, and $N_r$ represent the number of particles in bins corresponding to $q$, $p$, and $r$, respectively. 
Proceeding as in sec.~\ref{sec:method}, the lowest order moments are found to be
\begin{align}
\llangle Q_n P_m \rrangle = &  \langle N_qN_p\rangle \llangle q_i^{n} p_j^{m}\rrangle,  \\ 
\llangle Q_nQ_mPo \rrangle = & \langle N_qN_p\rangle \llangle q_i^{n+m} p_j^o\rrangle 
+ \langle N_q(N_q-1)N_p\rangle \llangle q_i^{n}q_j^{m} p_k^o\rrangle,  \\ 
\llangle Q_nQ_mQ_oP_p \rrangle =& \langle N_qN_p\rangle \llangle q_i^{n+m+o} p_j^p\rrangle 
  + \langle N_q(N_q-1)N_p\rangle \llangle q_i^{n+m}q_j^{o} p_k^p\rrangle 
  + \langle N_q(N_q-1)N_p\rangle \llangle q_i^{n+o}q_j^{m} p_k^p\rrangle   \\ \nonumber
  &\langle N_q(N_q-1)N_p\rangle \llangle q_i^{n}q_j^{m+o} p_k^p\rrangle   
  + \langle N_q(N_q-1)(N_q-2)N_p\rangle \llangle q_i^{n}q_j^{m}q_k^{o} p_l^{p}\rrangle,   \\
\llangle Q_nQ_mP_oP_p \rrangle =& \langle N_qN_p\rangle \llangle q_i^{n+m} p_j^{o+p}\rrangle 
+ \langle N_qN_p(N_p-1)\rangle \llangle q_i^{n+m}p_j^{o} p_k^p\rrangle 
+ \langle N_q(N_q-1)N_p \llangle q_i^{n}q_j^{m} p_k^{o+p}\rrangle   \\ \nonumber
&+ \langle N_q(N_q-1)N_p(N_p-1) \llangle \llangle q_i^{n}q_j^{m}p_k^{o} p_l^{p}\rrangle,    \\
\llangle Q_n P_m R_o \rrangle =& \langle N_qN_pN_r\rangle \llangle q_i^{n} p_j^{m} r_k^o\rrangle, \\ 
\llangle Q_n Q_m P_o R_p \rrangle =& \langle N_qN_pN_r\rangle \llangle q_i^{n+m} p_j^{o} r_l^p\rrangle
 + \langle N_q(N_q-1)N_pN_r\rangle \llangle q_i^{n} q_j^{m} p_k^{o} r_l^p\rrangle.
\end{align}
Based on the above expressions, one can iteratively obtain expressions for the moments $\llangle q_i^{n} q_j^{m} p_k^{o} r_l^p\rrangle$ in terms of moments of products of $Q$s, $P$s, and $R$s. Given the sum over $q$ and $p$ factorize, the moments become simple combinations of $Q$s and $P$s, and one gets at lowest orders
\begin{align}
\langle N_qN_p\rangle\llangle q_1^{n} p_1^{m}\rrangle =&  \llangle Q_n P_m \rrangle,  \\ 
\langle N_q(N_q-1)N_p\rangle  \llangle q_1^{n_1}q_2^{n_2} p_1^m\rrangle =& 
\llangle Q_{n_1}Q_{n_2}P_{m} \rrangle - \llangle Q_{n_1+n_2} P_m \rrangle \\
\langle N_q(N_q-1)(N_q-2)N_p\rangle  \llangle
\llangle q_1^{n_1}q_2^{n_2}q_3^{n_3} p_1^{m}\rrangle =&
\llangle Q_{n_1}Q_{n_2}Q_{n_3}P_{m} \rrangle - \sum_{(3)}\llangle Q_{n_1+n_2}Q_{n_3}P_{m} \rrangle \\ \nonumber 
&-2 \llangle Q_{n_1+n_2+n_3} P_m \rrangle. \\
\langle N_q \cdots(N_q-2)N_p(N_p-1)\rangle  \llangle
\llangle q_1^{n_1}q_2^{n_2}q_3^{n_3} p_1^{m_1}p_2^{m_2}\rrangle =&
-2 \llangle Q_{n1+n2+n3}P_{m1+m2}\rrangle 
+2 \llangle Q_{n1+n2+n3}P_{m1}P_{m2}\rrangle  \\ \nonumber 
&+\llangle Q_{n1+n2}Q_{n3}P_{m1+m2}\rrangle 
-\llangle Q_{n1+n2}Q_{n3}P_{m1}P_{m2}\rrangle  \\ \nonumber 
&+\llangle Q_{n1+n3}Q_{n2}P_{m1+m2}\rrangle 
-\llangle Q_{n1+n3}Q_{n2}P_{m1}P_{m2}\rrangle  \\ \nonumber 
&+\llangle Q_{n2+n3}Q_{n1}P_{m1+m2}\rrangle  
-\llangle Q_{n2+n3}Q_{n1}P_{m1}P_{m2}\rrangle \\ \nonumber 
&-\llangle Q_{n1}Q_{n2}Q_{n3}P_{m1+m2}\rrangle 
+\llangle Q_{n1}Q_{n2}Q_{n3}P_{m1}P_{m2}\rrangle.
\end{align}
Higher orders are increasingly tedious to compute for large values of $n$, $m$, and $o$. Fortunately, the scripts created for the computation of $\llangle q_1^{n_1} \, \cdots\, q_m^{n_m}\rrangle$ are trivially extendable to multiple variables provided one appropriately  considers all permutations of factors in $q$, $p$, and $r$. Finally, setting 
all exponents to unity one gets expressions of the form
\begin{align}
\langle N_qN_p\rangle\llangle q_1 p_1\rrangle =& \llangle Q_{1}P_{1}\rrangle,  \\ 
\langle N_q(N_q-1)N_p\rangle  \llangle q_1q_2p_1\rrangle =& 
-\llangle Q_{2}P_{1}\rrangle +\llangle Q_{1}^{2}P_{1}\rrangle\\
\langle N_q(N_q-1)(N_q-2)N_p\rangle   
\llangle q_1^{n_1}q_2^{n_2}q_3^{n_3} p_1^{m}\rrangle =&
+2 \llangle Q_{3}P_{1}\rrangle -3 \llangle Q_{2}Q_{1}P_{1}\rrangle +\llangle Q_{1}^{3}P_{1}\rrangle \\
\langle N_q \cdots(N_q-2)N_p(N_p-1)\rangle  
\llangle q_1^{n_1}q_2^{n_2}q_3^{n_3} p_1^{m_1}p_2^{m_2}\rrangle =&
-2 \llangle Q_{3}P_{2}\rrangle 
+2 \llangle Q_{3}P_{1}^{2}\rrangle \\ \nonumber 
&+3 \llangle Q_{2}Q_{1}P_{2}\rrangle 
-3 \llangle Q_{2}Q_{1}P_{1}^{2}\rrangle \\ \nonumber 
&-\llangle Q_{1}^{3}P_{2}\rrangle 
+\llangle Q_{1}^{3}P_{1}^{2}\rrangle,
\end{align}
and so on.

\section{Computation of Multi-particle Balance Functions}
\label{sec:Multi-particleBalanceFunctions}

Balance function of arbitrary  order $n$ can be defined using the procedure introduced in sec.~\ref{sec:Balance Functions} based on  differential correlators of the form $\langle \Delta q_1 \cdots \Delta q_n\rangle$
and corresponding $n$-order differential cumulant expansions. By construction, in the presence of $n$-particle correlations, the defined balance functions must yield unity when integrated over all particle transverse momenta ($p_{\rm T}>0$), azimuths, and rapidity. As in the case of the second and fourth orders, 
$n$-cumulant can be expanded into their $n$-tuplet charge combinations. Such 
decompositions are straightforwardly obtained by considering all ways to cluster $n$ particles into subgroups of $k\le n$ positively and $n-k$ negatively charged particles. Such decompositions are herewith denoted $k(+)n-k(-)$ in which $k$ and $n-k$ respectively represent the number of positively and negatively charged particles in a  decomposition of $n$ particle. Given the order in which the $+$ve and $-$ve particles are listed is inconsequential, the number of equivalent permutations 
is given by binomial coefficients 
\begin{align}
    \binom{n}{k}  = \frac{n!}{k! \left(n-k \right)!}.
\end{align}
Additionally, the sign of each term evidently depends on the number of negative particles in a particular decomposition.
Cumulants of order $n$ can thus be written 
\begin{align}
\label{eq:CnDecompoisition}
C_n(\vec p_1,\dots,\vec p_n) =& \sum_{k=0}^n (-1)^{n-k}  \binom{n}{k}
C_n^{k(+)n-k(-)}(\vec p_1,\dots,\vec p_6),
\end{align}

We split the cumulant decompositions 
to yield expressions of balance function corresponding to $m(+)$s given $m(-)$s and conversely, $m(-)$s given $m(+)$, for $m=n/2$.  This requires an additional factor of two  in the BF definitions. To avoid unnecessary  repetitions of labels $+$ and $-$ as in sec.~\ref{sec:Balance Functions}, we introduce the notations $B_n^{+-}$, with $n$ being even integers 2, 4, 6, etc,   
to indicate the  balance functions of $n/2$ positively charged particles found at momenta $\vec p_1, \ldots, \vec p_{n/2}$ given $n/2$ negatively charged particles are detected at  $\vec p_{n/2+1}, \ldots, \vec p_{n}$, and conversely, 
$B_n^{-+}$ shall indicate the BF of $n/2$ negatively charged particles found at momenta $\vec p_1, \ldots, \vec p_{n/2}$ given $n/2$ positively charged particles are detected at  $\vec p_{n/2+1}, \ldots, \vec p_{n}$.
The five lowest orders are thus written:
\begin{align}
\label{eq:B2+-}
B_2^{+-}(\vec p_1,\vec p_2) =& \frac{C_2^{+-}(\vec p_1,\vec p_2) -C_2^{--}(\vec p_1,\vec p_2)}{\langle N^{-} \rangle}, \\
\label{eq:B4+-}
B_4^{+-}(\vec p_1,\ldots,\vec p_4) =& \frac{1}{6} \times
\frac{
3C_4^{2+2-}
-4C_4^{1+3-}-C_4^{4-}
}{\langle N^{-}\left( N^{-}-1\right)\rangle }, \\
\label{eq:B6+-}
B_6^{+-}(\vec p_1,\dots,\vec p_6) =& \frac{1}{60} \times  \frac{10
 C_6^{3+3-} - 15 C_4^{2+4-} + 6 C_4^{1+5-} + C_4^{6-}}
 {\langle N^{-}\left( N^{-}-1\right)\left( N^{-}-2\right) \rangle}  \\
 \label{eq:B8+-}
B_8^{+-}(\vec p_1,\dots,\vec p_8) =&  \frac{1}{840}\times
\frac{
 35C_8^{4+4-} 
 - 56 C_8^{3+5-}
 + 25 C_8^{2+6-} 
  - 8 C_8^{1+7-} 
 + C_8^{8-} }
  {\langle N^{-}\left( N^{-}-1\right)\left( N^{-}-2\right)\left( N^{-}-3\right)\left( N^{-}-3\right)\rangle} \\ 
 \label{eq:B10+-}
B_{10}^{+-}(\vec p_1,\dots,\vec p_{10}) =&  \frac{1}{15120}\times
\frac{
 126C_{10}^{5+5-} 
 - 210C_{10}^{4+6-}
 + 120 C_{10}^{3+7-} 
 - 45 C_{10}^{2+8-} 
 + 10C_{10}^{1+9-} 
 - C_{10}^{10-} }
   {\langle N^{-}\left( N^{-}-1\right)\left( N^{-}-2\right)\left( N^{-}-3\right) \left( N^{-}-3\right)\left( N^{-}-4\right)\rangle}
\end{align}

Higher orders, $n>10$, are readily obtained based on Eq.~(\ref{eq:CnDecompoisition}) and can be written
\begin{align}
   \label{eq:Bn+-}
B_{n}^{+-}(\vec p_1,\dots,\vec p_{n}) =& \frac{(-1)^{n/2}}{N(n)} 
\sum_{k=0}^{n} (-1)^{n-k} \left(\frac{1}{2}\right)^{\delta_{n,2k}} \binom{n}{k} 
  C_{n}^{k(+)n-k(-)},
\end{align} 
where the normalization coefficient $N(n)$ is calculated according to 
\begin{align}
    N(n) = \frac{1}{2} \frac{n!}{(n/2)!}\langle N^{-}\left( N^{-}-1\right) \,\cdots\, \left( N^{-}-n+1\right)\rangle,
\end{align}
and $\delta_{n,2k}=1$ for $n=2k$ but otherwise vanishes.

\bibliography{main}

\end{document}